\input{header-1-packages}

\newcommand{\ungroup}[1]{#1}
\newcommand{\withbreak}[1]{\expandafter\ungroup#1}

\def\dounbracket[#1]{#1}

\def\ket#1{\left|#1\right\rangle}

\renewcommand\vec\mathbf

\newcommand\redsout{\bgroup\markoverwith{\textcolor{red}{\rule[0.5ex]{2pt}{1.4pt}}}\ULon}

\newcommand\dootimesall[2]{\ifx0#1\else\mathbf{#1}\ifx0#2\else\def\mytmp{\otimes\dootimesall{#2}}\expandafter\expandafter\expandafter\mytmp\fi\fi}

\usepackage{relsize}

\usepackage[acronyms, nohypertypes={acronym}, nopostdot, style=super, nonumberlist, toc]{glossaries}
\setacronymstyle{long-sc-short}
\glsenableentrycount

\newacronym{PNA}{pna}{particle-number-algorithm}
\newacronym{SFA}{sfa}{spin-flip-algorithm}

\newacronym{WF}{wf}{Wilson-Fisher}
\newacronym{AF}{af}{asymptotically free}

\newacronym{RG}{rg}{renormalization group}

\newacronym{QIS}{qis}{Quantum Information Science}
\newacronym{PPT}{ppt}{positive-semidefinite partial transpose}
\newacronym{KS}{ks}{Kogut-Susskind}
\newacronym{NPT}{npt}{negative partial transpose}
\newacronym{SMG}{smg}{symmetric mass generation}
\newacronym{SMF}{smf}{symmetric massive fermion}
\newacronym{MF}{mf}{massless fermion}
\newacronym{SSB}{ssb}{spontaneous symmetry breaking}

\newacronym{AS}{as}{Anti-Symmetric}

\newacronym[longplural={conformal field theories}]{CFT}{cft}{conformal field theory}
\newacronym[longplural={lattice field theories}]{LFT}{lft}{lattice field theory}
\newacronym[longplural={effective field theories}]{EFT}{eft}{effective field theory}
\newacronym[longplural={quantum field theories}]{QFT}{qft}{quantum field theory}
\newacronym[longplural={lattice gauge theories}]{LGT}{lgt}{lattice gauge theory}
\newacronym[longplural={monomer-dimer tensor-networks}]{MDTN}{mdtn}{monomer-dimer tensor-network}

\newacronym[]{DMRG}{dmrg}{Density Matrix Renormalization Group}
\newacronym[]{TFIM}{tfim}{Transverse Field Ising Model}

\newacronym[]{LOCC}{locc}{Local Operations and Classical Communicaton}
\newacronym[]{OBC}{obc}{open boundary conditions}

\newacronym{MPS}{mps}{matrix product states}

\newacronym{JLP}{jlp}{Jordan-Lee-Preskill}

\newacronym{BBN}{bbn}{big bang nucleosynthesis}

\newacronym{LEC}{lec}{low-energy constant}

\newacronym{QCD}{qcd}{quantum chromodynamics}
\newacronym{MC}{mc}{Monte Carlo}

\newacronym{IR}{ir}{infrared}
\newacronym{UV}{uv}{ultraviolet}

\newacronym{QED}{qed}{quantum electrodynamics}
\newacronym{SNR}{snr}{signal-to-noise ratio}

\newacronym{NLSM}{nlsm}{nonlinear sigma model}

\newacronym{CL}{cl}{Complex Langevin}

\newacronym{CSA}{csa}{Cartan subalgebra}

\newacronym{AFQMC}{afqmc}{auxiliary field quantum Monte Carlo}
\newacronym{iHMC}{ihmc}{imaginary-mass Hybrid Monte Carlo}

\newacronym{MCMC}{mcmc}{Markov Chain Monte Carlo}

\newacronym{QI}{qi}{quantum information}

\newacronym{irrep}{{\rm irrep}}{irreducible representation}

\newacronym{ASQR}{asqr}{antisymmetric qubit regularization}

\begin{document}

\title{Frustration-Induced Expressibility Limitations in Variational Quantum Algorithms}
\author{Sandip Maiti\,\orcidlink{0000-0002-5248-5316}}
\email{sandipmaiti73@gmail.com \\ sandip.maiti@ccnu.edu.cn}
\affiliation{Key Laboratory of Quark and Lepton Physics (MOE) and Institute of Particle Physics, Central China Normal University, Wuhan 430079, China}

\date{\today}

\begin{abstract}
Geometric frustration, arising from competing interactions that prevent simultaneous energy minimization, presents a fundamental challenge for variational quantum algorithms applied to quantum many-body systems. We investigate the transverse-field Ising model on a square lattice with frustrated diagonal coupling and show that geometric frustration leads to strongly inhomogeneous correlations that are difficult to capture using standard Hamiltonian-inspired ansätze with global parameters. As a result, the required circuit depth increases significantly in the intermediate-field regime. We demonstrate that this limitation is not caused by optimization difficulties such as barren plateaus, but instead arises from insufficient expressibility of the ansatz. By introducing bond-resolved variational parameters, we recover accurate results at reduced circuit depth. We also study low-energy excitations and find that near-degenerate spectra in the frustrated regime further challenge variational methods. Our results provide a clear physical explanation for the limitations of variational quantum algorithms in frustrated systems and suggest improved ansatz design strategies for quantum simulation.
\end{abstract}

\maketitle

\section{Introduction}

Variational quantum algorithms have emerged as leading candidates for near-term quantum simulation of strongly correlated many-body systems, particularly on noisy intermediate-scale quantum (NISQ) devices \cite{Preskill:2018jim}. Among these, the Variational Quantum Eigensolver (VQE) \cite{Peruzzo:2013bzg, Tilly:2021jem} and the Quantum Approximate Optimization Algorithm (QAOA) \cite{Farhi:2014ych, Zhou:2018fwi} have been widely applied to quantum spin models and optimization problems. Recent studies have demonstrated that these approaches can capture ground-state properties of frustrated systems on both classical simulators and quantum hardware \cite{Lotshaw:2022kpa, Lively:2024ijv, Mohtashim:2021sgm, Kirmani:2025pvb}. However, their performance degrades significantly in frustrated regimes, often requiring increased circuit depth and yielding reduced fidelity.

Despite these advances, a key challenge remains: the underlying physical origin of this degradation is not well understood, and systematic strategies to overcome these limitations are largely unexplored. In addition, most existing work focuses on ground-state properties, leaving the behavior of variational methods for low-energy excitations in frustrated systems less understood.

Frustrated quantum spin systems provide a natural setting to investigate these questions. Competing interactions give rise to highly nontrivial energy landscapes and densely packed low-energy manifolds, leading to strong quantum fluctuations and complex correlation structures. In particular, frustration can induce spatially inhomogeneous correlations even in translationally invariant systems, posing a fundamental challenge for Hamiltonian-inspired variational ansätze that employ globally shared parameters. Complex energy landscapes in spin systems are also known to pose challenges for alternative quantum paradigms such as quantum annealing \cite{King:2018rbd}.

In this work, we investigate these issues using the transverse-field Ising model (TFIM) on the square lattice with frustrated diagonal couplings. This model provides a controlled platform in which geometric frustration induces heterogeneous bond correlations and nontrivial spectral structure, while remaining amenable to exact diagonalization (ED) for moderate system sizes.

We show that geometric frustration leads to strongly inhomogeneous correlations that cannot be captured efficiently by variational ansätze with globally shared parameters, such as the Hamiltonian Variational Ansatz (HVA). As a result, the required circuit depth increases sharply in the frustrated regime, and performance saturates at fixed depth. Using gradient-norm analysis, we demonstrate that this degradation is not associated with barren plateaus, but instead reflects an intrinsic expressibility limitation of the variational manifold.

To overcome this limitation, we introduce a symmetry-informed bond-resolved HVA, in which independent parameters are assigned to each interaction term. This construction enables the circuit to adapt to local correlation structure and significantly improves accuracy at reduced circuit depth.

Beyond ground-state properties, we investigate low-energy excitations using symmetry-resolved variational circuits and Variational Quantum Deflation (VQD) \cite{Higgott:2018doo}. While excitation gaps are accurately captured in regimes with well-separated spectra, near-degenerate manifolds in the frustrated regime present additional challenges for variational approaches. Extending our analysis to larger system sizes using Krylov subspace methods, we show that these limitations persist beyond system sizes accessible to ED.

In this work, we establish a direct connection between geometric frustration, correlation structure, and variational expressibility. This perspective provides general design principles for constructing problem-informed ansätze capable of efficiently representing complex many-body states, with implications for quantum simulation of frustrated systems.

The rest of this paper is organized as follows. In Sec.~\ref{model}, we introduce the model and discuss its physical properties. In Sec.~\ref{sec:variational}, we describe the variational ansätze and algorithms used in this work. Sec.~\ref{sec:numerics} outlines the numerical methods and observables. In Sec.~\ref{sec:results_ground}, we present our main results, including ground-state benchmarking, expressibility analysis, and excitation gaps. Finally, we summarize our findings in Sec.~\ref{conclusion}.

\section{Model and Phase Structure}
\label{model}

We consider a TFIM on a two-dimensional square lattice with both nearest-neighbor and diagonal couplings of equal strength ($J_1 = J_2 = J$),
\begin{equation}
H = - J \sum_{\langle ij \rangle} Z_i Z_j 
    - J \sum_{\langle\langle ij \rangle\rangle} Z_i Z_j 
    - h \sum_{i} X_i,
\label{eq:tfim}
\end{equation}
where $Z_i$ and $X_i$ are Pauli operators acting on site $i$, $\langle ij \rangle$ and $\langle\langle ij \rangle\rangle$ denote nearest-neighbor and diagonal (next-nearest-neighbor) pairs, respectively, $J$ is the Ising coupling strength, and $h$ is the transverse field. Throughout this work we focus on the antiferromagnetic case ($J<0$), where the presence of triangular plaquettes induces geometric frustration.

\subsection{Classical Limit and Geometric Frustration}

In the classical limit ($h=0$), antiferromagnetic interactions on lattices containing triangular plaquettes lead to geometric frustration, as it is impossible to simultaneously satisfy all pairwise interactions. This phenomenon is well established in the triangular-lattice Ising model, where the ground state is macroscopically degenerate, with a finite residual entropy density \cite{PhysRev.79.357}.

In the present model, triangular frustration is induced on a square lattice through diagonal couplings. Each elementary triangle contains three bonds, leading to competing constraints that cannot be simultaneously satisfied. As a result, the system exhibits a highly degenerate manifold of low-energy configurations. Minimizing the energy locally favors two spins aligned antiparallel and one parallel on each triangle, producing a complex and highly nontrivial energy landscape that suppresses simple N\'eel order.

\subsection{Quantum Fluctuations and Order-by-Disorder}

Introducing a transverse field ($h>0$) generates quantum fluctuations that lift the classical degeneracy via an order-by-disorder mechanism. In the case of the nearest-neighbor triangular-lattice TFIM, this leads to the emergence of a three-sublattice ordered phase with a $\sqrt{3}\times\sqrt{3}$ structure at small fields \cite{PhysRevLett.84.4457, PhysRevB.63.224401}. In this regime, long-range order arises from quantum fluctuations acting on the extensively degenerate classical manifold.

In the frustrated geometry considered here, a similarly nontrivial ordered phase is expected to emerge at small transverse fields. While the underlying lattice geometry differs from the triangular lattice, the presence of frustrated triangular plaquettes leads to competing interactions and complex correlation patterns, which can give rise to enlarged unit cells and multi-sublattice ordering.

As $h$ increases, quantum fluctuations compete more strongly with the interaction energy. For the nearest-neighbor triangular lattice with periodic boundary conditions, the system undergoes a quantum phase transition to a paramagnetic phase at a critical field $h_c/|J| \approx 1.65$ \cite{PhysRevB.68.104409}. In finite systems, this transition is replaced by a crossover, characterized by a minimum in the excitation gap and enhanced low-energy spectral density near the critical region. A qualitatively similar crossover behavior is expected in the present model, although the precise critical properties depend on lattice geometry and boundary conditions.

\subsection{Paramagnetic Regime}

In the paramagnetic regime (PM), where the transverse field dominates ($h \gg |J|$), the ground state approaches a product state
\begin{equation}
|\psi_{\mathrm{PM}}\rangle = \bigotimes_i |+\rangle_i,
\end{equation}
where $|+\rangle$ is the eigenstate of $X$ with eigenvalue $+1$. In this regime, the excitation gap increases with $h$ and correlations become weak, making the system efficiently representable by shallow variational circuits.

\subsection{Spectral Structure and Finite-Size Effects}

For finite system sizes, signatures of the thermodynamic transition persist as:
\begin{itemize}
    \item a minimum in the excitation gap,
    \item enhanced susceptibility,
    \item increased entanglement entropy,
    \item clustering of low-lying eigenvalues.
\end{itemize}

In the frustrated low-field regime, remnants of the classical degeneracy lead to near-degenerate low-energy states and small excitation gaps that decrease with increasing system size. This spectral crowding plays a central role in variational simulations, as resolving closely spaced eigenstates requires increased circuit expressibility.

\subsection{Comparison with the Ferromagnetic Case}

For comparison, the ferromagnetic case ($J>0$) on the same lattice is unfrustrated and belongs to the (2+1)-dimensional Ising universality class. The classical ground state is nondegenerate and exhibits simple long-range order at $h=0$, and the quantum phase transition occurs at a significantly larger critical field ($h_c/|J| \approx 4.7$) \cite{Hamer:2006zz}. 

This contrast highlights the role of geometric frustration in generating a highly degenerate classical manifold and a more complex low-energy structure in the antiferromagnetic case.

In this work, we focus on representative field values $h=0.5, 1.0, 1.5$, and $4.0$, which probe the frustrated regime, crossover region, and paramagnetic phase.

\subsection{Inhomogeneous Correlations}

\begin{figure}
\centering
\includegraphics[width=0.22\textwidth]{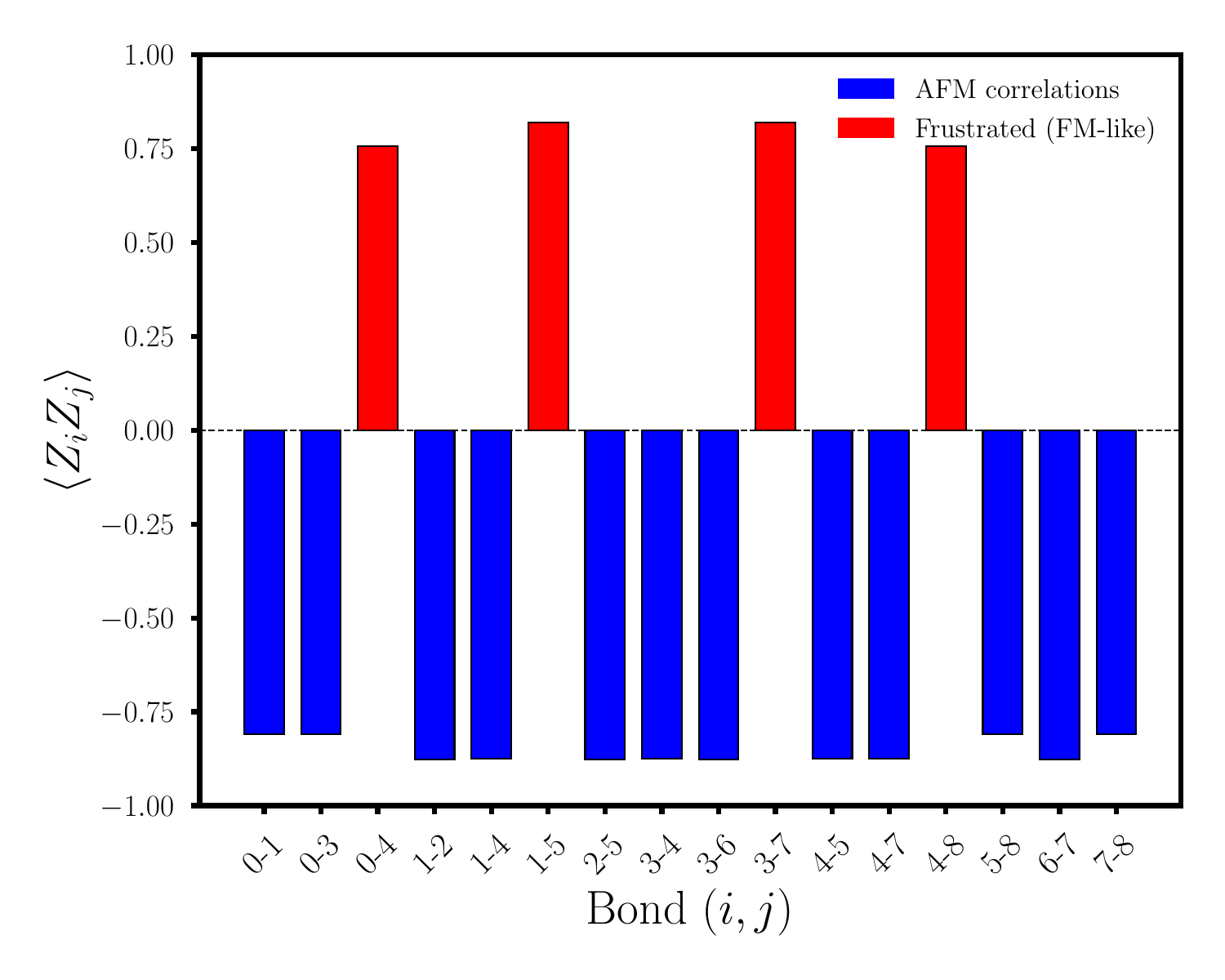}
\includegraphics[width=0.234\textwidth]{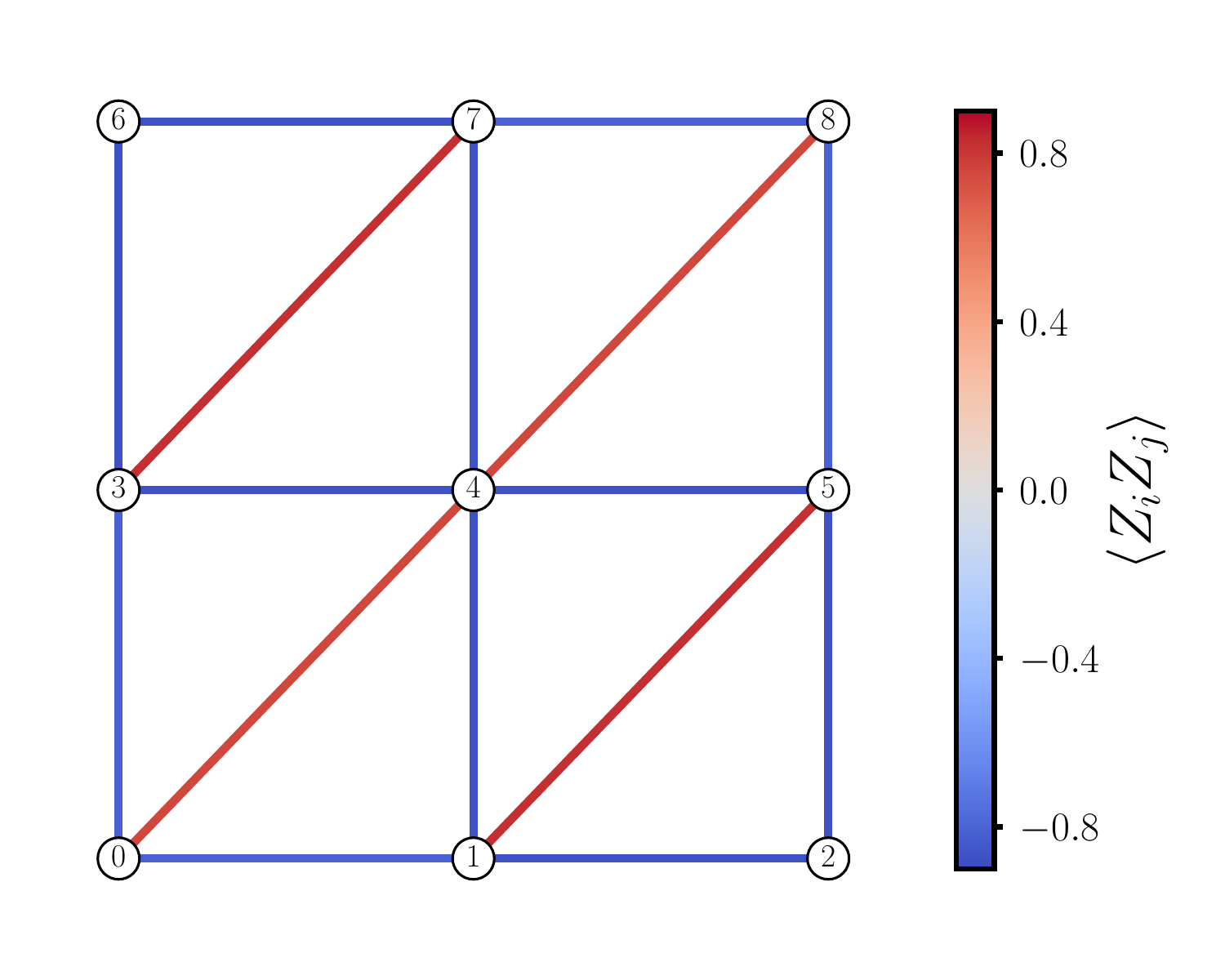}
\caption{Bond-resolved two-site correlations $\langle Z_i Z_j\rangle$ for the $3\times3$ square lattice with frustrated diagonal couplings at $h=0.5$ from ED. In each triangle, two bonds are energetically satisfied (antiferromagnetic), while one bond remains frustrated. This leads to a heterogeneous pattern of correlations across the lattice, with diagonal bonds typically exhibiting weaker or positive correlations.}
\label{fig:ED-corr}
\end{figure}

A key consequence of geometric frustration is the emergence of spatially inhomogeneous correlations, even in translationally invariant Hamiltonians. To illustrate this, we compute bond-resolved two-site correlations $\langle Z_i Z_j \rangle$ using exact diagonalization (ED) for the $3\times3$ lattice at $h=0.5$, shown in Fig.~\ref{fig:ED-corr}.

The results reveal a clear separation of bonds into distinct classes corresponding to satisfied and frustrated interactions. This bond-dependent structure indicates that the ground state cannot be described by a uniform correlation profile, but instead requires spatially resolved variational flexibility. This observation directly motivates the bond-resolved variational ansatz introduced in the next section.

\section{Variational Quantum Algorithms}
\label{sec:variational}

To investigate the ability of near-term quantum circuits to capture frustrated quantum magnetism, we employ variational quantum algorithms within a classical optimization loop. In this section, we define the Hamiltonian Variational Ansatz, the bond-resolved HVA, and a hardware-efficient ansatz, and outline the approaches used to access excited states, including Variational Quantum Deflation and a symmetry-resolved formulation of the bond-resolved ansatz.

\subsection{Variational Quantum Eigensolver Framework}

The Variational Quantum Eigensolver (VQE) \cite{Peruzzo:2013bzg} seeks to approximate the ground state of a Hamiltonian $H$ by minimizing the energy expectation value
\begin{equation}
E(\boldsymbol{\theta}) = 
\langle \psi(\boldsymbol{\theta}) | H | \psi(\boldsymbol{\theta}) \rangle,
\end{equation}
over a parameterized family of states $|\psi(\boldsymbol{\theta})\rangle$. 
The variational parameters $\boldsymbol{\theta}$ are updated by a classical optimizer until convergence.

In this work, all simulations are performed using exact statevector simulation, allowing direct computation of energies, fidelities, and gradient norms.

\subsection{Hamiltonian Variational Ansatz}

Our primary ansatz is a Hamiltonian-inspired circuit structurally equivalent to the QAOA \cite{Farhi:2014ych} applied to the TFIM.

\begin{figure}
\centering
\includegraphics[width=0.47\textwidth]{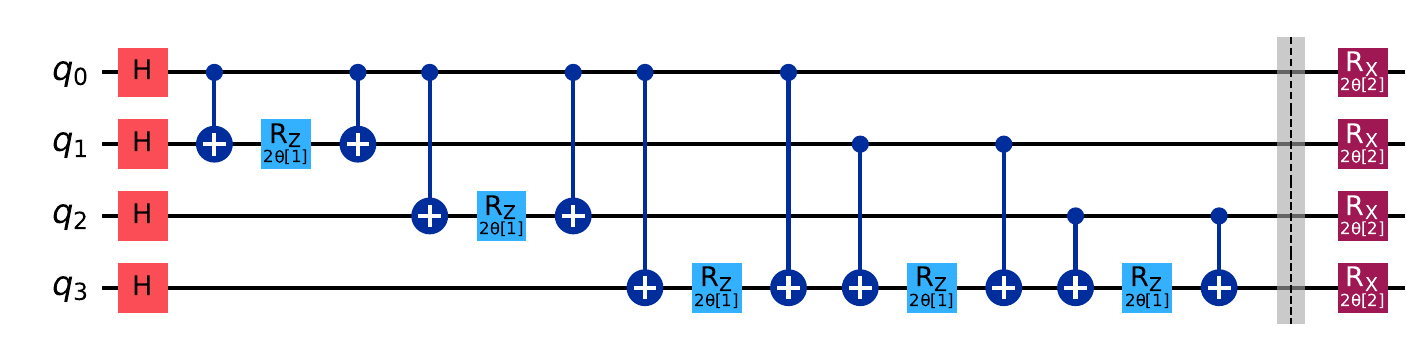}
\caption{Hamiltonian Variational Ansatz (HVA) circuit for $N=4$ qubits with a single layer. The initial state is $\ket{+}^{\otimes N}$. In each layer, all $ZZ$ interactions share a single parameter $\theta[1]$, followed by a global $X$-rotation mixer with parameter $\theta[2]$. This yields a total of $2p$ variational parameters for $p$ layers.}
\label{fig:FS-hva}
\end{figure}

The ansatz for the ground state consists of $p$ alternating layers of Ising interaction and transverse-field evolution,
\begin{equation}
|\psi(\boldsymbol{\gamma},\boldsymbol{\beta})\rangle = \prod_{k=1}^{p} e^{-i \beta_k H_X}
e^{-i \gamma_k H_{ZZ}} |+\rangle^{\otimes N},
\label{eq:hva}
\end{equation}
where
\begin{align}
H_{ZZ} &= -J \sum_{\langle ij \rangle} Z_i Z_j, \\
H_X &= - \sum_i X_i.
\end{align}

Each layer is parameterized by two real parameters $(\gamma_k, \beta_k)$, yielding a total of $2p$ variational parameters. 
The initial state $|+\rangle^{\otimes N}$ corresponds to the ground state of $H_X$ in the large-field limit. The structure of the HVA circuit is illustrated in \cref{fig:FS-hva}.

This ansatz preserves the structure of the problem Hamiltonian and can be interpreted as a variational Trotterization of real-time evolution. Its structured form reduces parameter count and aligns the variational manifold with physically relevant directions in Hilbert space.

\subsection{Bond-Resolved Hamiltonian Variational Ansatz}

As demonstrated in Fig.~\ref{fig:ED-corr}, geometric frustration induces strongly inhomogeneous correlations, even in small system sizes. This directly challenges Hamiltonian-inspired ansätze with globally shared parameters, which implicitly assume a uniform correlation structure across all bonds.

To address this limitation, we introduce a bond-resolved HVA in which independent parameters are assigned to each interaction term.

\begin{figure}
\centering
\includegraphics[width=0.47\textwidth]{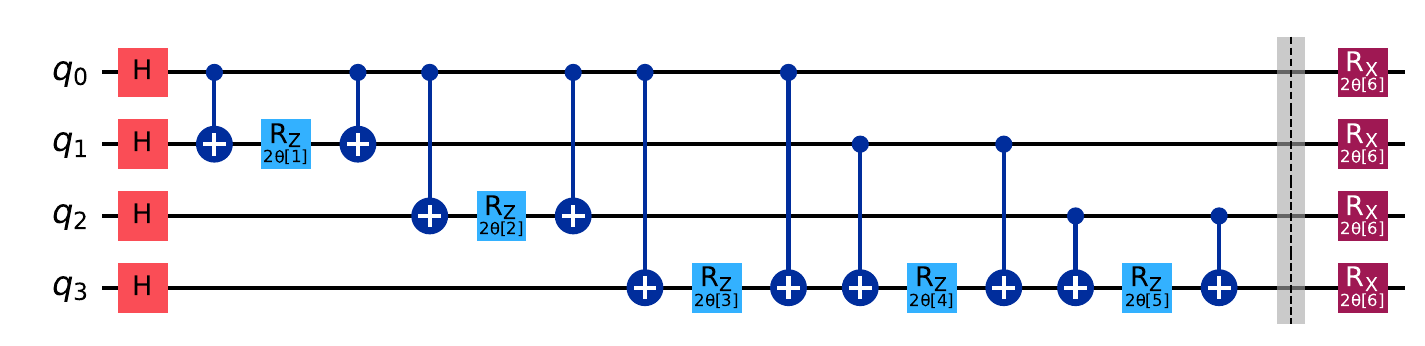}
\caption{Bond-resolved HVA circuit for $N=4$ qubits (single layer). The initial state is $\ket{+}^{\otimes N}$. Each layer consists of $ZZ$ interactions on all edges with independent parameters $\theta[k]$, followed by a global $X$-rotation mixer, resulting in $p(|\mathcal{E}|+1)$ variational parameters.}
\label{fig:FS-bhva}
\end{figure}

In the standard HVA, all Ising interactions share a single parameter $\gamma_k$ per layer, enforcing uniform evolution under $H_{ZZ}$. However, in frustrated systems, the local correlation structure varies significantly from bond to bond. Capturing such heterogeneous correlations requires additional variational flexibility.

We therefore generalize the ansatz by introducing bond-dependent parameters $\gamma_{ij}^{(k)}$ for each edge $\langle ij \rangle$. The resulting variational state is
\begin{equation}
|\psi(\boldsymbol{\gamma}, \boldsymbol{\beta})\rangle = \prod_{k=1}^{p} e^{-i \beta_k H_X}
\left(\prod_{\langle ij \rangle} e^{-i \gamma_{ij}^{(k)} Z_i Z_j} \right) |+\rangle^{\otimes N}.
\end{equation}

This bond-resolved ansatz increases the number of variational parameters per layer from two to $|\mathcal{E}| + 1$, where $|\mathcal{E}|$ is the number of edges in the lattice. The additional degrees of freedom enable the circuit to adapt independently to each bond, thereby capturing the spatially inhomogeneous correlation patterns induced by geometric frustration. A circuit illustration is shown in \cref{fig:FS-bhva}.

\begin{figure}
\centering
\includegraphics[width=0.235\textwidth]{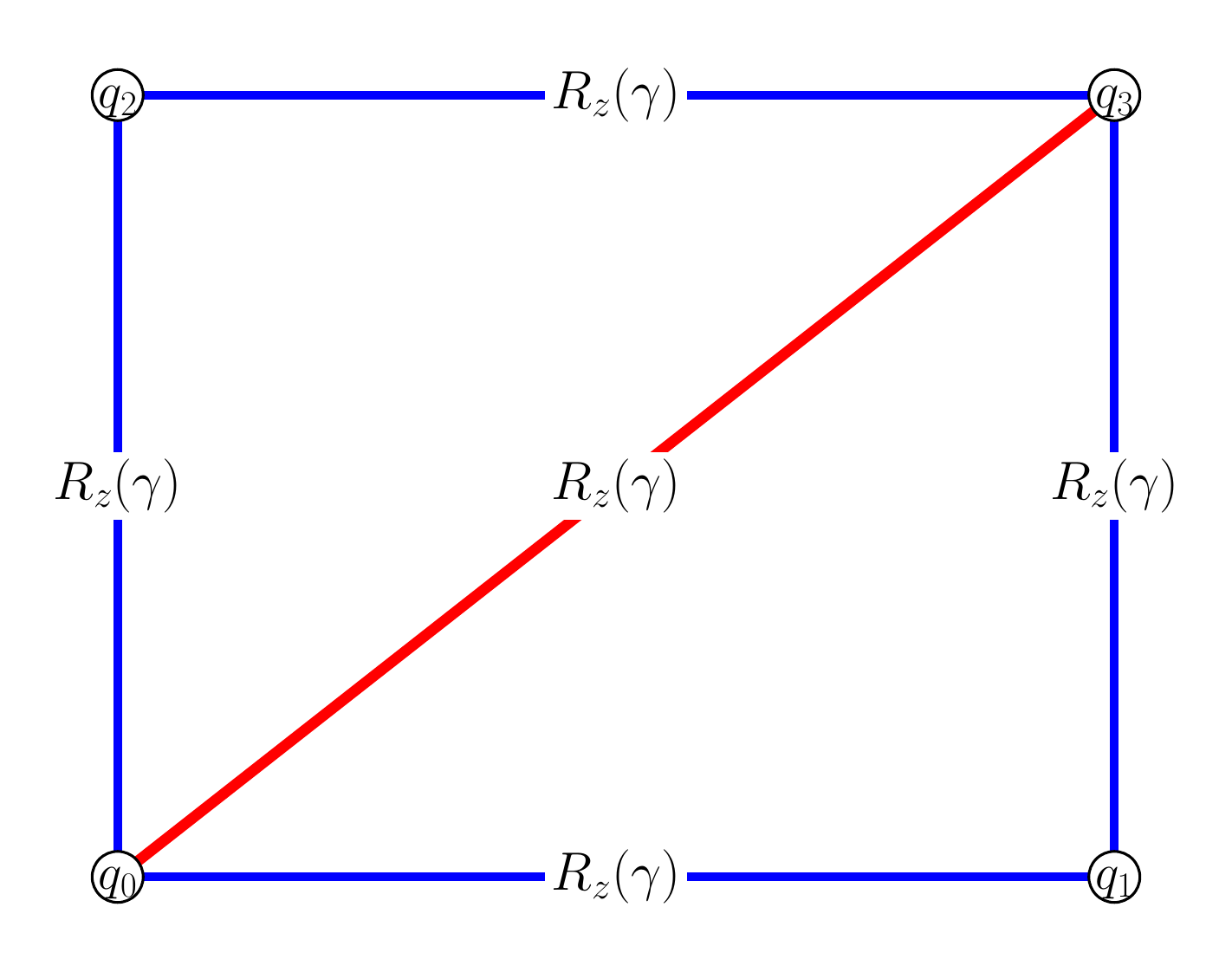}
\includegraphics[width=0.235\textwidth]{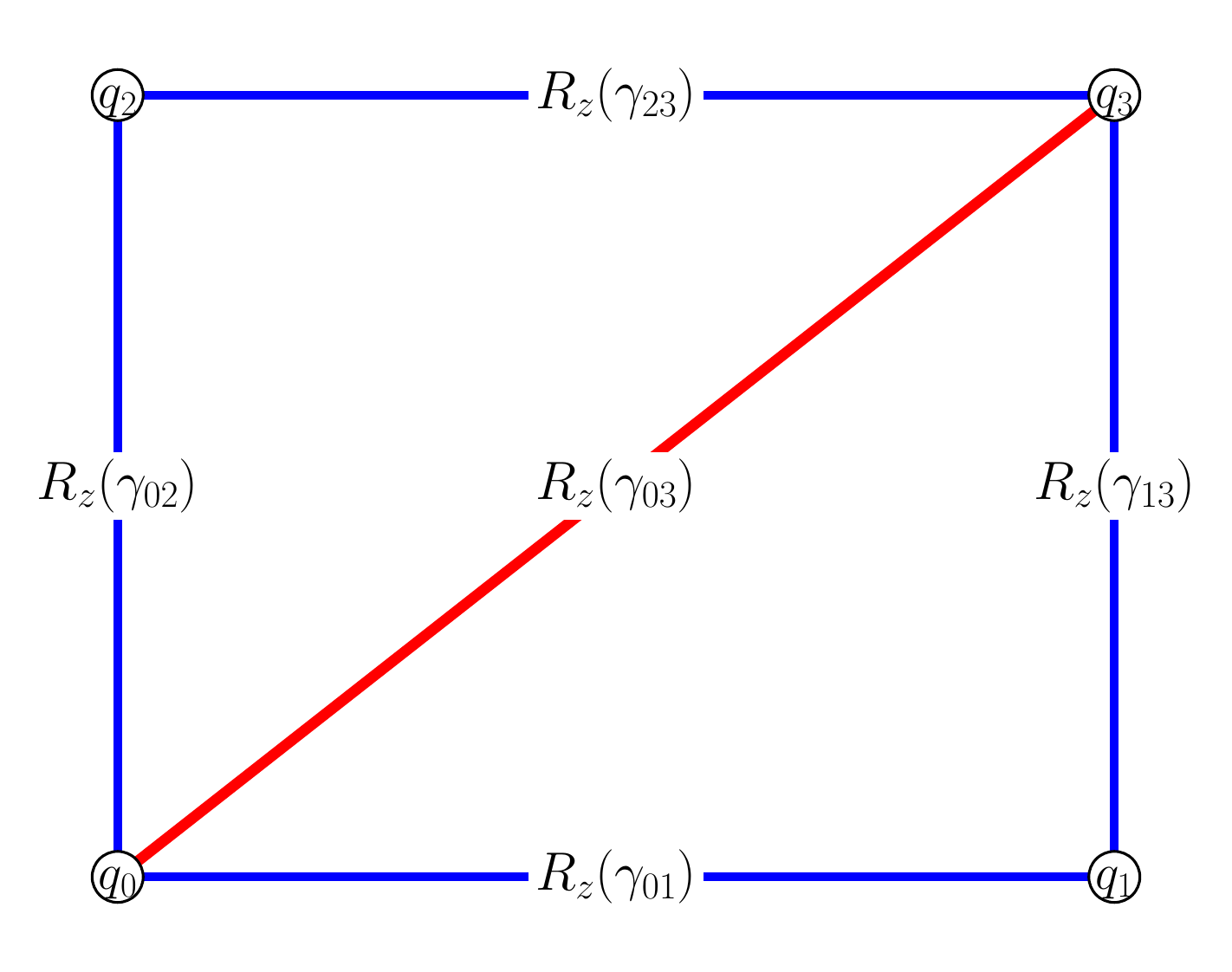}
\caption{Lattice representation of the variational ansatz on the $2\times 2$ system, where each lattice site corresponds to a qubit ($N=4$). (Left) HVA with a single global parameter $\gamma$ applied uniformly across all bonds. (Right) Bond-resolved HVA with independent parameters $\gamma_{ij}$ on each bond.}
\label{fig:corr-circuit}
\end{figure}

In contrast to the standard HVA, which constrains all bonds to evolve identically, the bond-resolved formulation allows for bond-dependent correlations, significantly enhancing expressibility in frustrated regimes. The figure \ref{fig:corr-circuit} illustrates the qubit–lattice mapping and the corresponding parameter assignments in both ansätze.

\subsection{Hardware-Efficient Ansatz}

To assess the role of Hamiltonian structure, we also consider a hardware-efficient ansatz (HEA) \cite{Kandala:2017vok, McClean:2015vup}. The ansatz consists of alternating layers of single-qubit rotations and entangling gates arranged in a fixed connectivity pattern. In this work, we employ rotation layers composed of $R_y$ and $R_z$ gates with linear nearest-neighbor entanglement.

Each layer applies
\begin{equation}
U_{\mathrm{HEA}}^{(k)} =
\left(\prod_{i=1}^{N} R_y(\theta_{i,k}^{(1)}) R_z(\theta_{i,k}^{(2)}) \right) U_{\mathrm{ent}},
\end{equation}
where $U_{\mathrm{ent}}$ denotes a fixed entangling pattern (e.g., linear nearest-neighbor CNOTs), and $\theta_{i,k}^{(1,2)}$ are independent variational parameters.

Unlike the HVA, the HEA does not encode Hamiltonian structure explicitly and typically contains a larger number of parameters per layer. This ansatz provides a benchmark for generic expressibility without physics-informed design.

\subsection{Variational Quantum Deflation}

To access excited states, we employ two complementary approaches. First, we use the Variational Quantum Deflation (VQD) method \cite{Higgott:2018doo}. After obtaining an approximate ground state $|\psi_0\rangle$, the excited state is found by minimizing a modified cost function,
\begin{equation}
C(\boldsymbol{\theta}) = \langle \psi(\boldsymbol{\theta}) | H | \psi(\boldsymbol{\theta}) \rangle + \beta \sum_{m < k} |\langle \psi(\boldsymbol{\theta}) | \psi_m \rangle|^2,
\label{eq:vqd}
\end{equation}
where $\{\psi_m\}$ are previously computed lower-energy states and $\beta > 0$ is a penalty coefficient. The penalty term suppresses overlap with known states, driving the optimization toward orthogonal eigenstates. In this work, we validate VQD against ED for small lattices and analyze its stability across different spectral regimes.

In addition to the penalty-based approach, we exploit a global $\mathbb{Z}_2$ symmetry of the TFIM to directly resolve low-energy states in different symmetry sectors. The Hamiltonian commutes with the operator $\prod_i X_i$, allowing eigenstates to be classified by their parity eigenvalue $\pm 1$.

In practice, this symmetry-resolved construction is implemented using the bond-resolved Hamiltonian variational ansatz, where the additional variational flexibility enables stable convergence within a fixed symmetry sector. For ground-state calculations, we initialize the variational circuit in the even-parity sector using the product state $|+\rangle^{\otimes N}$, which satisfies $\prod_i X_i |+\rangle^{\otimes N} = + |+\rangle^{\otimes N}$.

To access the first excited state, we initialize the circuit in the odd-parity sector by preparing a state with eigenvalue $-1$ under $\prod_i X_i$, obtained by applying a single local spin flip to $|+\rangle^{\otimes N}$. While a translation-invariant superposition could also be used, this simpler initialization is sufficient to restrict the optimization to the desired symmetry sector and requires a lower circuit depth.

Since the variational circuit preserves this symmetry, the optimization remains confined within the chosen sector, allowing independent approximation of the lowest-energy states in each symmetry sector. This symmetry-resolved approach complements VQD and provides a robust framework for extracting excitation energies, particularly in regimes with near-degenerate spectra where penalty-based methods become less stable, consistent with previous symmetry-resolved variational studies \cite{Maiti:2024jwk}.

\subsection{Gradient Analysis}

To characterize optimization landscapes, we compute the Euclidean norm of the energy gradient,
\begin{equation}
\|\nabla_{\boldsymbol{\theta}} E\| = \left(\sum_i \left| \frac{\partial E}{\partial \theta_i} \right|^2 \right)^{1/2}.
\end{equation}

Gradients are evaluated using the parameter-shift rule \cite{Schuld:2018uel}, which is exact for rotation generators with two eigenvalues. Monitoring the gradient norm during optimization allows us to distinguish between expressibility-limited convergence and barren plateau behavior. In particular, vanishing gradients at initialization would signal a barren plateau, whereas finite residual gradients indicate convergence within a restricted variational manifold.

\section{Numerical Setup and Observables}
\label{sec:numerics}

\subsection{System Sizes and Parameters}
We study the antiferromagnetic ($J=-1$) TFIM on square lattices with frustrated diagonal couplings, of size $L\times L$ with open boundary conditions, for $L=2,3,4$. The total number of spins is $N = L^2$, corresponding to a Hilbert-space dimension of $2^N$. Open boundary conditions are chosen to minimize circuit depth and gate count, consistent with typical NISQ implementations. For benchmarking, we compute exact eigenvalues and eigenvectors using sparse ED for all system sizes considered.

To probe distinct physical regimes, we focus on representative transverse-field strengths
\begin{equation}
h = 0.5,\; 1.0,\; 1.5,\; 4.0,
\end{equation}
which correspond to the frustrated regime ($h=0.5$), the crossover and near-critical region ($h=1.0,\,1.5$), and the paramagnetic phase ($h=4.0$). For $L=3$, we additionally perform a fine-grained scan over a wider range of $h$ to resolve crossover behavior (see Appendix \ref{app:observables}).

\subsection{Krylov Subspace Diagonalization}

Exact diagonalization becomes computationally prohibitive for the $5\times5$ lattice ($N=25$ spins), where the Hilbert-space dimension is $2^{25}$. To estimate low-energy eigenvalues for this system, we employ a Krylov subspace (Rayleigh--Ritz) diagonalization approach inspired by the Subspace Krylov Quantum Diagonalization (SKQD) framework~\cite{yu2025}.

Starting from one or more reference states $\{|\phi_a\rangle\}$, we construct a Krylov subspace of dimension $D$,
\begin{equation}
\mathcal{K}_D = \mathrm{span}\{|\phi_a\rangle, H|\phi_a\rangle, H^2|\phi_a\rangle, \ldots, H^{D-1}|\phi_a\rangle\}.
\end{equation}
In practice, the basis vectors are generated by repeated application of the Hamiltonian followed by orthogonalization to ensure linear independence.

Within this subspace, we compute the projected Hamiltonian and overlap matrices,
\begin{align}
H_{ij} &= \langle \phi_i | H | \phi_j \rangle, \\
S_{ij} &= \langle \phi_i | \phi_j \rangle,
\end{align}
and solve the generalized eigenvalue problem
\begin{equation}
H c = E S c .
\end{equation}
The resulting eigenvalues provide variational approximations to the low-energy spectrum, with the lowest eigenvalue yielding an estimate of the ground-state energy $E_0$. We apply this method to the $5\times5$ lattice to obtain ground-state energy estimates beyond the reach of exact diagonalization.

\subsection{Optimization Details}

Variational parameters are optimized using gradient-based classical optimizers with fixed maximum iteration counts. For the standard HVA, we employ the L-BFGS-B optimizer, while for the bond-resolved HVA and the HEA, we use SLSQP, which we find provides more stable and efficient convergence in higher-dimensional parameter spaces.

For the HVA, each circuit of depth $p$ contains $2p$ parameters $(\gamma_k,\beta_k)$, while the bond-resolved ansatz involves a larger number of parameters proportional to the number of edges. Unless otherwise stated, we fix a reference depth $p=8$ for systematic comparison across lattice sizes and field values. In regimes where fidelity saturation occurs at fixed depth (notably for $h=0.5$), we perform additional depth-scaling studies.

All simulations are performed using exact statevector evaluation, allowing direct computation of expectation values and overlaps without sampling noise. A detailed comparison of optimizer performance is provided in Appendix~\ref{app:optimizer}.

\subsection{Physical Observables}

To characterize both physical and algorithmic performance, we compute the following observables:

\paragraph{Ground-State Energy Error.}
\begin{equation}
\Delta E_0 = E_0^{\mathrm{VQE}} - E_0^{\mathrm{ED}},
\end{equation}
which measures variational accuracy relative to ED.

\paragraph{State Fidelity.}
\begin{equation}
f = |\langle \psi_{\mathrm{VQE}} | \psi_{\mathrm{ED}} \rangle|^2,
\end{equation}
which quantifies the overlap with the corresponding exact eigenstate obtained from ED.

\paragraph{Excitation Gap.}
Using variational methods, we extract the first excited-state energy and define the excitation gap as
\begin{equation}
\Delta = E_1 - E_0.
\end{equation}
Gap estimates are compared directly with ED results.

\paragraph{Susceptibility.}
We compute the magnetic susceptibility via
\begin{equation}
\chi = \frac{1}{N} \left( \langle M_z^2 \rangle - \langle M_z \rangle^2 \right),
\end{equation}
where $N$ is the total number of spins and $M_z = \sum_i Z_i$.

\paragraph{Entanglement Entropy.}
We compute bipartite von Neumann entropy
\begin{equation}
S = -\mathrm{Tr}(\rho_A \log_2 \rho_A),
\end{equation}
where $\rho_A$ is the reduced density matrix of a subsystem obtained by tracing out the complement.

\paragraph{Gradient Norm.}
To analyze optimization landscapes, we monitor
\begin{equation}
\|\nabla_{\boldsymbol{\theta}} E\|,
\end{equation}
evaluated using the parameter-shift rule.

Together, these observables allow us to assess both physical fidelity and algorithmic performance across regimes and system sizes.

\section{Variational Benchmarking of Ground and Low-Energy States}
\label{sec:results_ground}

\subsection{Exact Spectral Structure}

We first analyze the exact low-energy spectrum obtained from ED. As expected for the frustrated square lattice, the excitation gap shows strong dependence on both field strength and system size. In the ordered regime ($h = 0.5$), the gap is very small and decreases significantly with increasing lattice size, reflecting remnants of the classically degenerate manifold and multiple closely spaced low-energy states. In the crossover regime ($h = 1.0$), the gap begins to open but remains strongly size-dependent, indicating persistent correlations. Near the critical region ($h \approx 1.5$), the gap increases while still showing moderate finite-size effects. In the paramagnetic regime ($h = 4.0$), the gap is large and only weakly dependent on system size, consistent with a gapped phase.

This spectral structure provides the baseline for interpreting variational algorithm performance.

\subsection{Variational Performance and Expressibility}

We now examine the performance of the HVA across different regimes at fixed circuit depth. In the paramagnetic regime (large $h$), the HVA achieves near-unity fidelity for all system sizes considered, with ground-state energy errors below $10^{-3}$. This confirms that shallow QAOA-like circuits efficiently represent weakly correlated product-like states.

In contrast, in the frustrated ordered regime (small $h$), performance degrades significantly as system size increases. While small systems are captured accurately, larger lattices exhibit saturation of fidelity at fixed depth. Increasing circuit depth improves accuracy only up to a point, beyond which gains become negligible.

To distinguish whether this degradation arises from insufficient depth or intrinsic ansatz limitations, we perform depth-scaling studies in the frustrated regime. We find that fidelity increases with circuit depth but eventually plateaus, even though gradients remain finite. This indicates that optimization is not hindered by barren plateaus, but rather constrained by the restricted variational manifold of the HVA.

These observations suggest that the degradation in performance is not solely due to limited circuit depth, but is instead linked to the increasing complexity of correlations in the frustrated regime.

\begin{figure}
\centering
\includegraphics[width=0.45\textwidth]{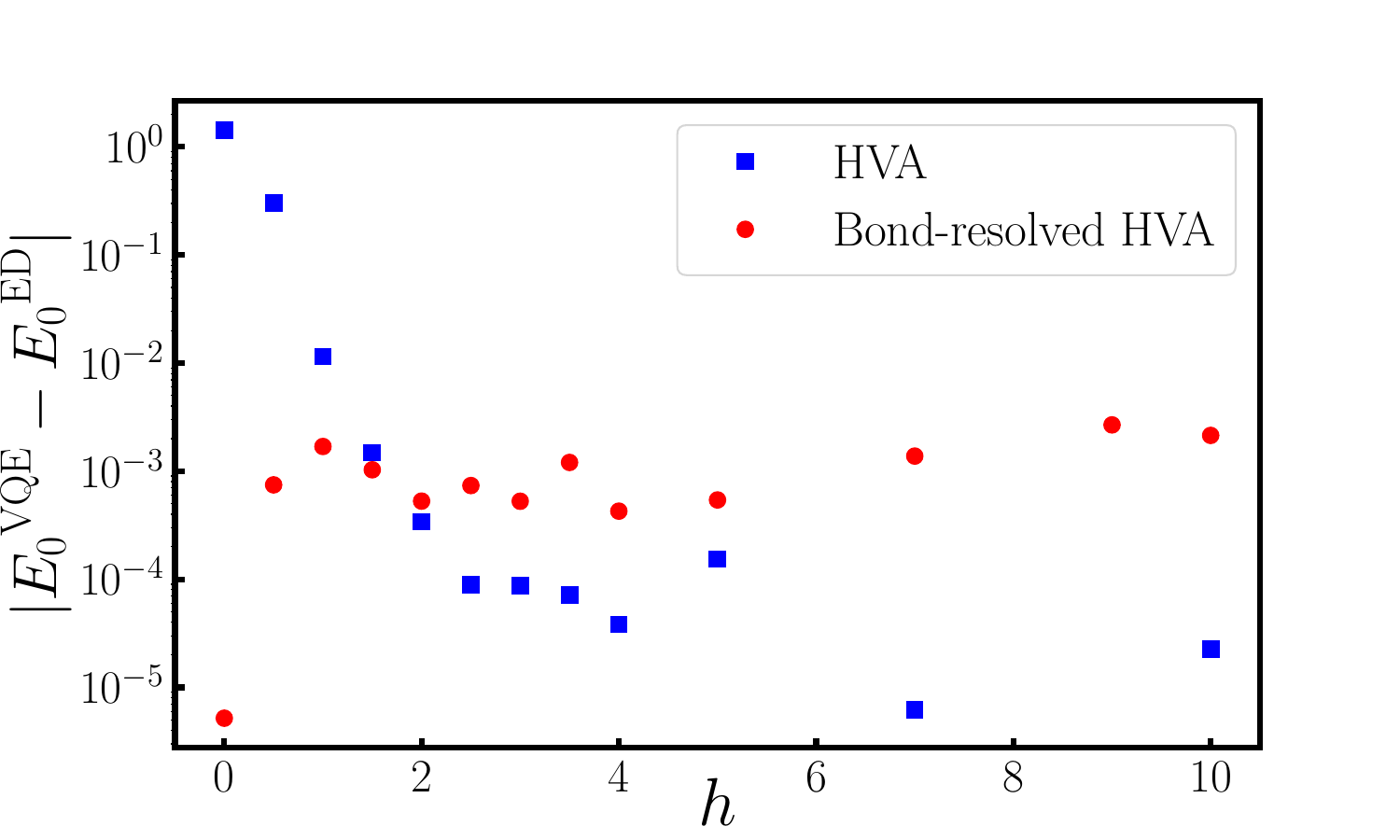}
\caption{Relative energy error for the ground state for the frustrated AFM on $3 \times 3$ lattice at fixed circuit depth $p=8$. The bond-resolved HVA yields lower error than the standard HVA, with a clear advantage in the frustrated regime (small $h$).}
\label{fig:E-err}
\end{figure}

Motivated by this behavior, we compare the ground-state energy error obtained using the standard HVA and the bond-resolved HVA for the $3\times 3$ lattice at fixed circuit depth $p=8$ across different transverse fields, as shown in Fig.~\ref{fig:E-err}. In the frustrated regime, the standard HVA exhibits relatively large errors that persist even at moderate circuit depth. In contrast, the bond-resolved ansatz significantly reduces the energy error, indicating a substantial improvement in representational capability. At larger $h$, both ansätze achieve comparable accuracy, consistent with the reduced correlation complexity in the paramagnetic regime.

\begin{figure}
\centering
\includegraphics[width=0.45\textwidth]{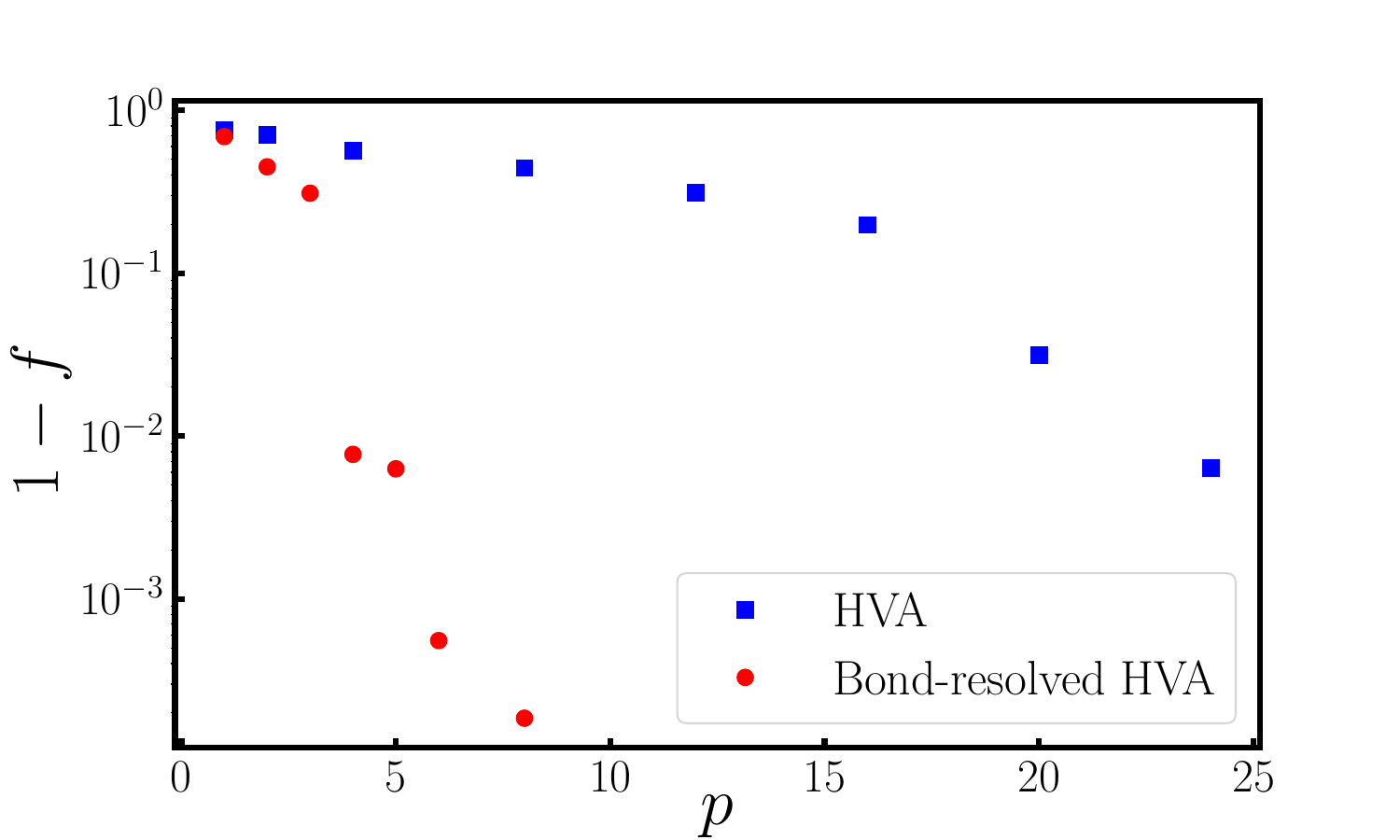}
\caption{Ground-state infidelity $(1 - f)$ as a function of circuit depth $p$ for the frustrated AFM on a $3 \times 3$ lattice at $h = 0.5$. The bond-resolved HVA achieves rapid convergence and attains high fidelity at shallow depths, whereas the standard HVA exhibits slower convergence and eventual saturation. At the maximum depth considered, the bond-resolved ansatz ($p = 8$) yields $E_0 \approx -8.7390$, in excellent agreement with ED ($E_0 \approx -8.7398$). In contrast, the standard HVA requires significantly greater depth ($p = 24$) and still shows a noticeable deviation ($E_0 \approx -8.7260$).}
\label{fig:fid-comp}
\end{figure}

To further understand this improvement, we analyze the depth dependence of the variational accuracy in the strongly frustrated regime ($h = 0.5$), shown in Fig.~\ref{fig:fid-comp}. For the standard HVA, the infidelity ($1-f$) decreases slowly with increasing circuit depth and exhibits clear saturation, reflecting limited expressibility. In contrast, the bond-resolved ansatz achieves a rapid reduction in infidelity and reaches significantly higher fidelity at much smaller depths.

This indicates that increasing circuit depth alone is insufficient when the variational structure is mismatched with the underlying correlation pattern, and that bond-resolved parameterization provides a more efficient representation.

\begin{figure}
\centering
\includegraphics[width=0.45\textwidth]{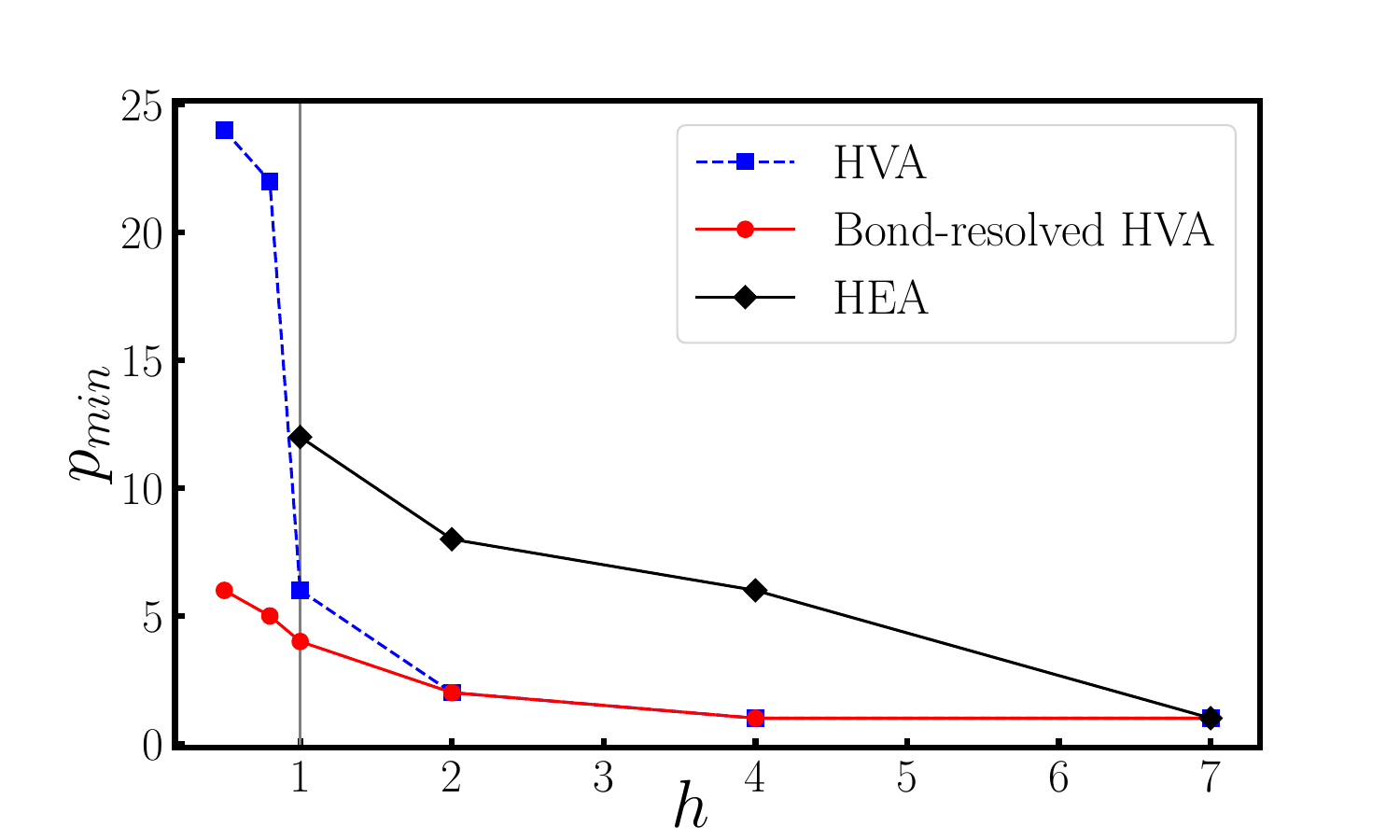}
\caption{Minimum circuit depth $p_{\mathrm{min}}$ required to reach fidelity $f>0.99$ for the ground state of the $3\times3$ lattice, as a function of $h$. The required depth increases significantly at low $h$ due to strong frustration, partially alleviated by bond-resolved ansätze. The vertical line at $h=1.0$ marks the crossover to the regime where shallow circuit depths suffice.}
\label{fig:depth-scaling-frus}
\end{figure}

We summarize this behavior by extracting the minimum circuit depth $p_{\min}$ required to reach high fidelity ($f > 0.99$) as a function of the transverse field, as shown in Fig.~\ref{fig:depth-scaling-frus}. In the frustrated regime, the required depth increases sharply for the standard HVA, reflecting the growing complexity of the ground-state correlations. The bond-resolved ansatz substantially reduces $p_{\min}$ across this regime, alleviating the expressibility bottleneck.

To further isolate the role of geometric frustration, we compare the depth requirements of the HVA on frustrated and non-frustrated square lattices at fixed system size. As shown in Appendix~\ref{app:depth-scaling}, using the standard HVA in both cases, the non-frustrated square lattice is accurately captured with shallow circuits across all $h$, whereas the frustrated square lattice requires substantially larger circuit depth to achieve high fidelity in the low-field regime.

To quantify the associated circuit cost, we report in Table~\ref{tab:cnot_scaling} the total number of CNOT gates required to achieve high fidelity ($f > 0.99$). The results show a pronounced increase in gate count in the frustrated regime for the standard HVA, reflecting the growing expressibility requirements. In contrast, the bond-resolved ansatz significantly reduces the required resources, indicating that incorporating local variational flexibility leads to more efficient circuit representations.

At larger $h$, both ansätze require only shallow circuits, consistent with the weakly correlated nature of the paramagnetic phase and the reduced complexity of the underlying quantum state.

\begin{figure}
\centering
\includegraphics[width=0.449\textwidth]{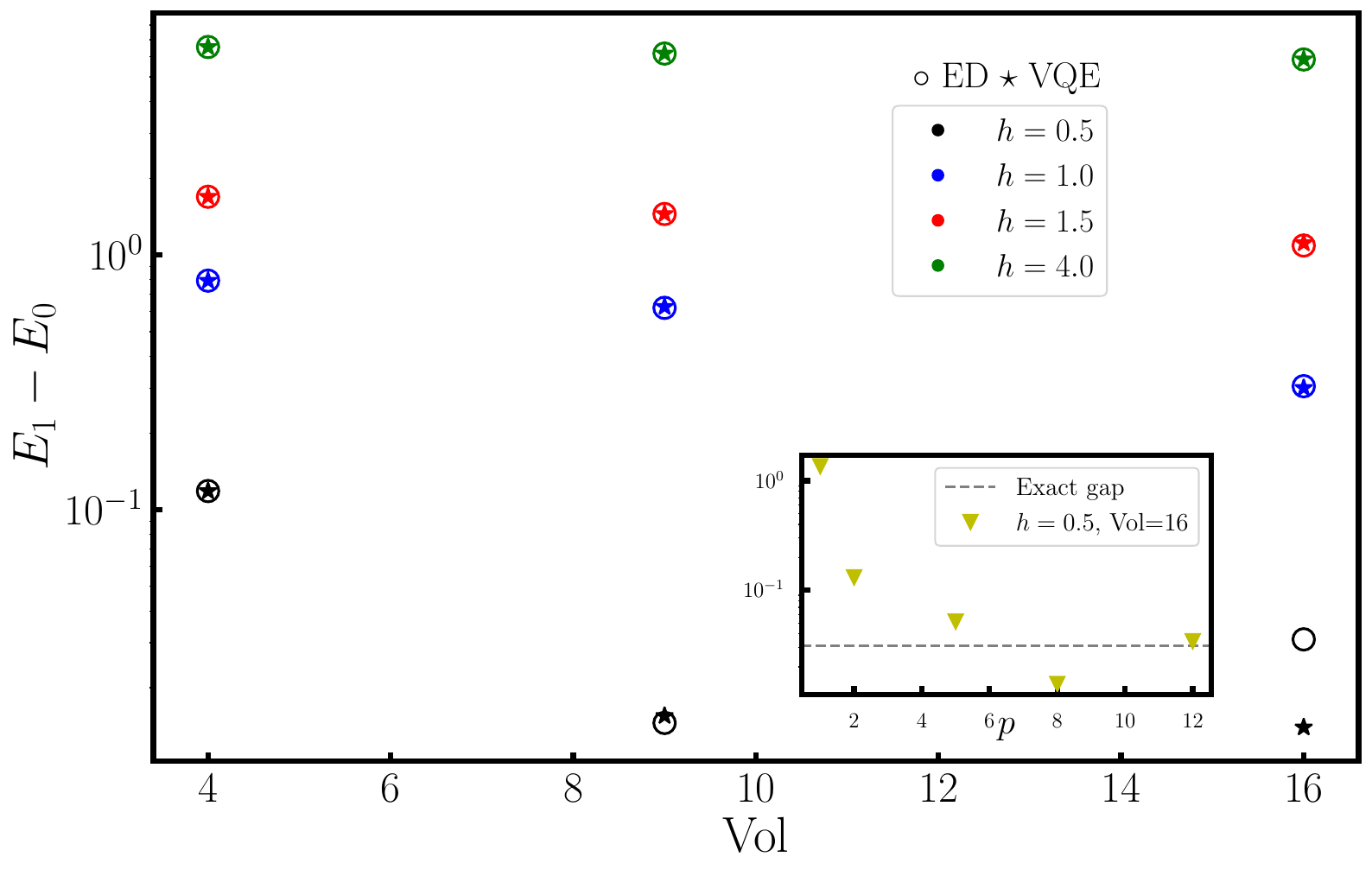}
\caption{Finite-size scaling of the excitation gap $\Delta = E_1 - E_0$ for the frustrated AFM at $h = 0.5, 1.0, 1.5,$ and $4.0$. Open circles denote ED results, while stars denote bond-resolved VQE results (using $\mathbb{Z}_2$ symmetry sectors) are compared with ED. A fixed circuit depth $p=8$ is used for all VQE calculations across system sizes and field values. The noticeable deviation for the $4\times 4$ lattice at $h=0.5$ arises from the limited circuit depth. The inset shows the convergence of the gap with increasing circuit depth $p$ for $\text{Vol}=16$ at $h=0.5$.}
\label{fig:AFM-gap}
\end{figure}

Finally, we assess the ability of the variational approach to capture low-energy spectral properties by comparing the excitation gap $\Delta = E_1 - E_0$ with ED, as shown in Fig.~\ref{fig:AFM-gap}. The ground and first excited states are obtained using the bond-resolved Hamiltonian variational ansatz within distinct symmetry sectors of the global $\mathbb{Z}_2$ operator $\prod_i X_i$, ensuring consistent identification of the lowest-energy states in each sector.

Across all field values, the variational results reproduce the qualitative behavior of the excitation gap, including its suppression in the frustrated regime and its growth in the paramagnetic phase. Quantitatively, good agreement is observed in regimes where the spectral gap is sufficiently large and the low-energy states are well separated.

In the strongly frustrated regime ($h = 0.5$), where the spectrum becomes dense and near-degenerate, deviations become more pronounced as system size increases. The inset in Fig.~\ref{fig:AFM-gap} shows the convergence of the lowest energy gap with increasing circuit depth $p$ for the $4\times 4$ lattice at $h=0.5$, demonstrating that the deviation in the main plot arises from the limited depth used.

\begin{figure}
\centering
\includegraphics[width=0.235\textwidth]{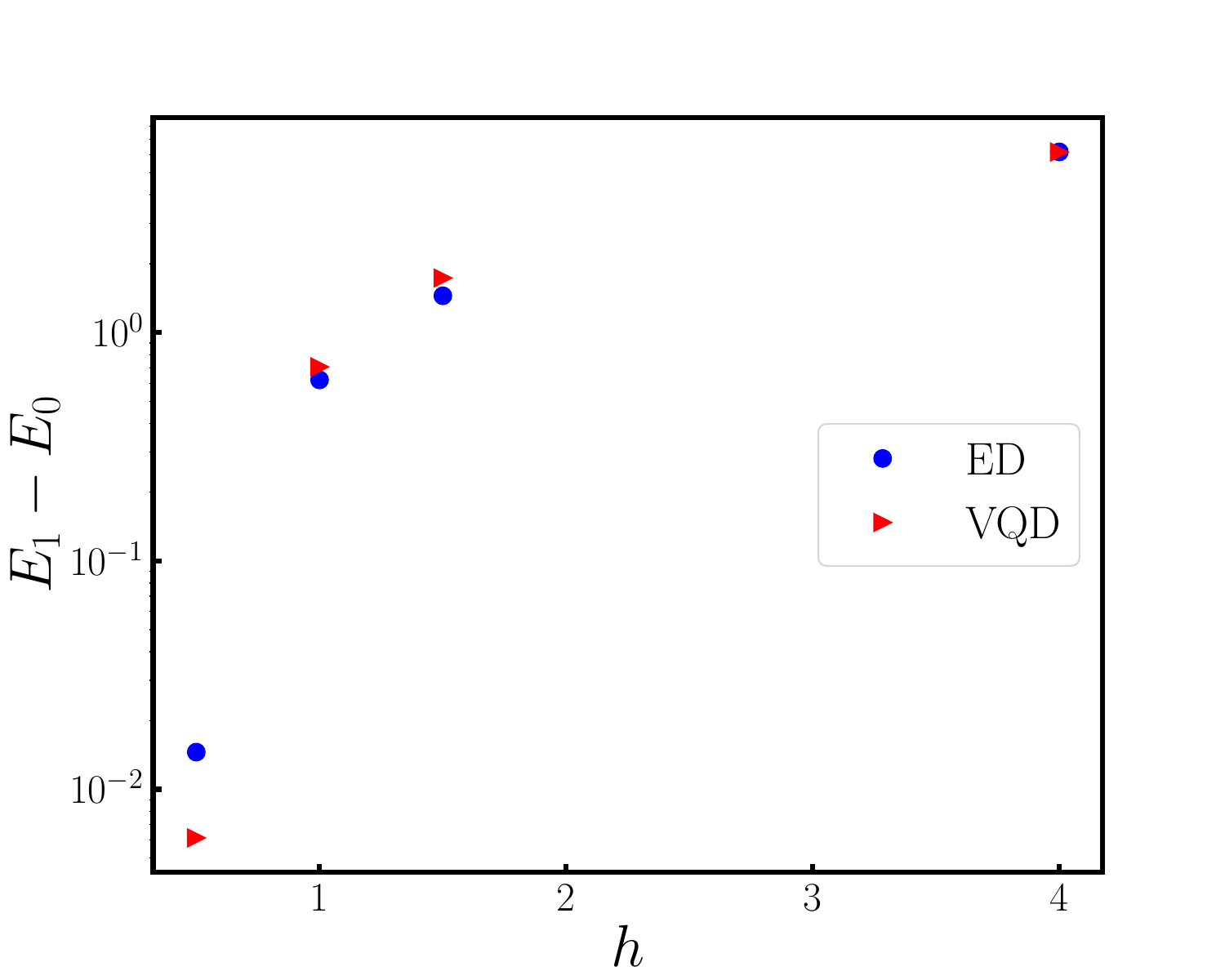}
\includegraphics[width=0.235\textwidth]{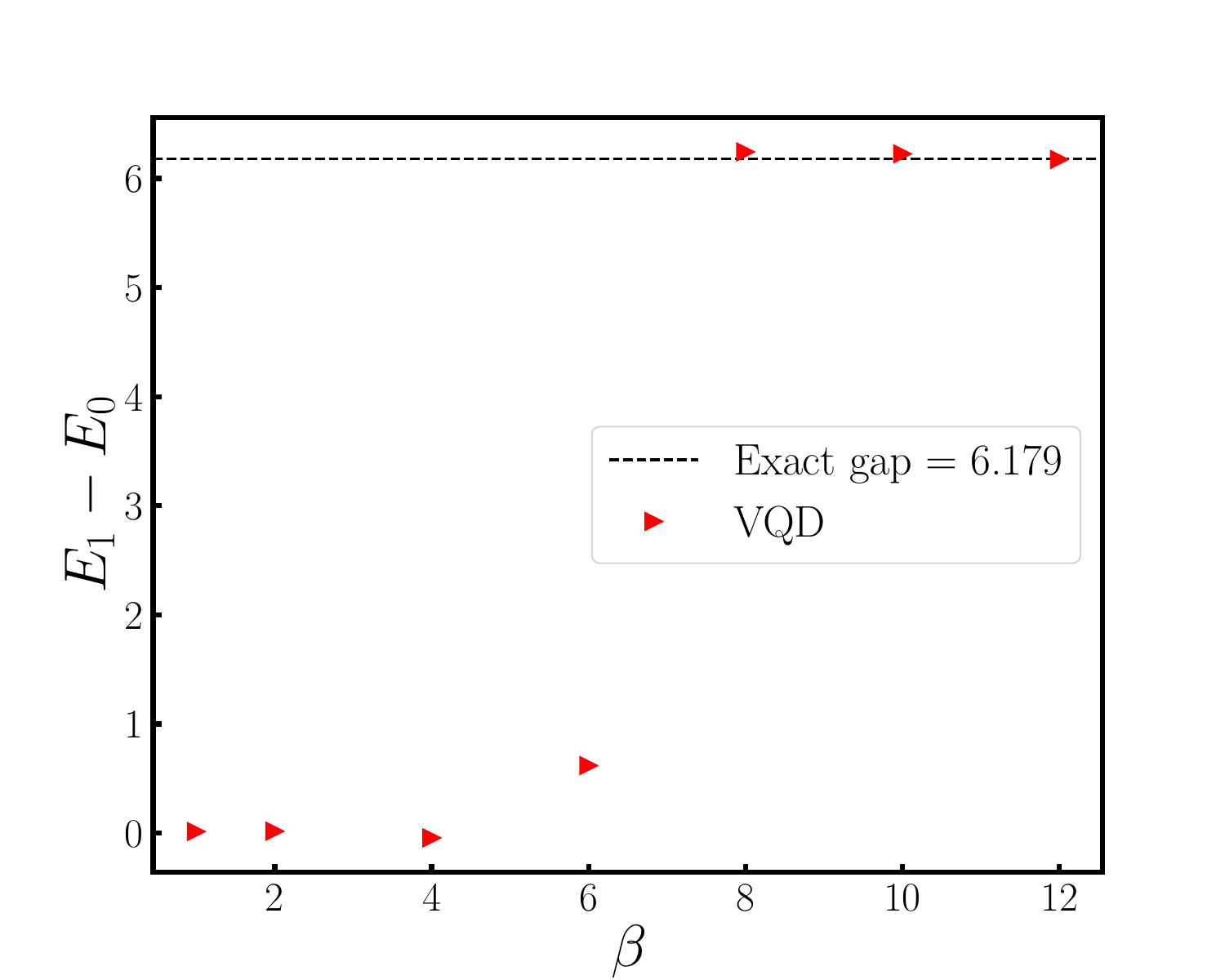}
\caption{Excitation gap $E_1 - E_0$ as a function of the transverse field $h$ for the $3\times 3$ frustrated square lattice. (Left) Gap as a function of $h$ from VQD ($p=8$, $\beta=12$) compared with ED. Good agreement is observed in the large-$h$ paramagnetic regime, where the spectrum is well separated. As $h$ decreases, deviations appear due to near-degenerate low-energy states. (Right) Gap as a function of $\beta$ at $h=4.0$, showing convergence once $\beta$ exceeds the exact gap.}
\label{fig:VQD-gap}
\end{figure}

To validate this approach, we benchmark excitation gaps obtained using VQD against ED (left panel of Fig.~\ref{fig:VQD-gap}). The choice of the penalty parameter $\beta$ is illustrated in the right panel. We find that VQD reproduces the gap accurately in regimes with well-separated spectra, but becomes less stable in the strongly frustrated regime due to the presence of near-degenerate low-energy states. This comparison highlights the advantage of symmetry-resolved variational approaches, which avoid overlap penalties and provide a more stable route to accessing low-energy excitations.

\begin{table*}
\renewcommand{\arraystretch}{1.6}
\setlength{\tabcolsep}{6pt}
\centering
\begin{tabular}{c|ccc|ccc|ccc}
\hline
 & \multicolumn{3}{c|}{$L=2$} & \multicolumn{3}{c|}{$L=3$} & \multicolumn{3}{c}{$L=4$} \\
$h$ & HVA & \makecell{Bond-resolved \\ HVA} & HEA & HVA & \makecell{Bond-resolved \\ HVA} & HEA & HVA & \makecell{Bond-resolved \\ HVA} & HEA \\
\hline
0.5 & 40 (p=4) & 20 (p=2) & 6 (p=2) & 768 (p=24) & 192 (p=6) & ...     & ...    & 792 (p=12)  & ... \\
1.0 & 30 (p=3) & 20 (p=2) & 6 (p=2) & 192 (p=6)  & 128 (p=4) & 96 (p=12)  & 924 (p=14) & 528 (p=8) & ... \\
1.5 & 20 (p=2) & 10 (p=1) & 6 (p=2) & 192 (p=6)  & 96 (p=3)  & 64 (p=8)   & 396 (p=6) & 330 (p=5) & ... \\
4.0 & 10 (p=1) & 10 (p=1) & 6 (p=2) & 32 (p=1)   & 32 (p=1)  & 48 (p=6)   & 66 (p=1)  & 66 (p=1)  & ... \\
\hline
\end{tabular}
\caption{Total CNOT count required to achieve ground-state fidelity $f > 0.99$ for different lattice sizes and transverse-field values. The minimum circuit depth $p$ required to reach this fidelity is indicated in parentheses, and the reported CNOT count corresponds to the full circuit at that depth. Data for the HEA at $L=4$ are not shown, as achieving $f>0.99$ would require substantially larger circuit depths and optimization resources than those considered in this work, and is further hindered by slow convergence in the high-dimensional parameter space.}
\label{tab:cnot_scaling}
\end{table*}

\begin{figure}
\centering
\includegraphics[width=0.233\textwidth]{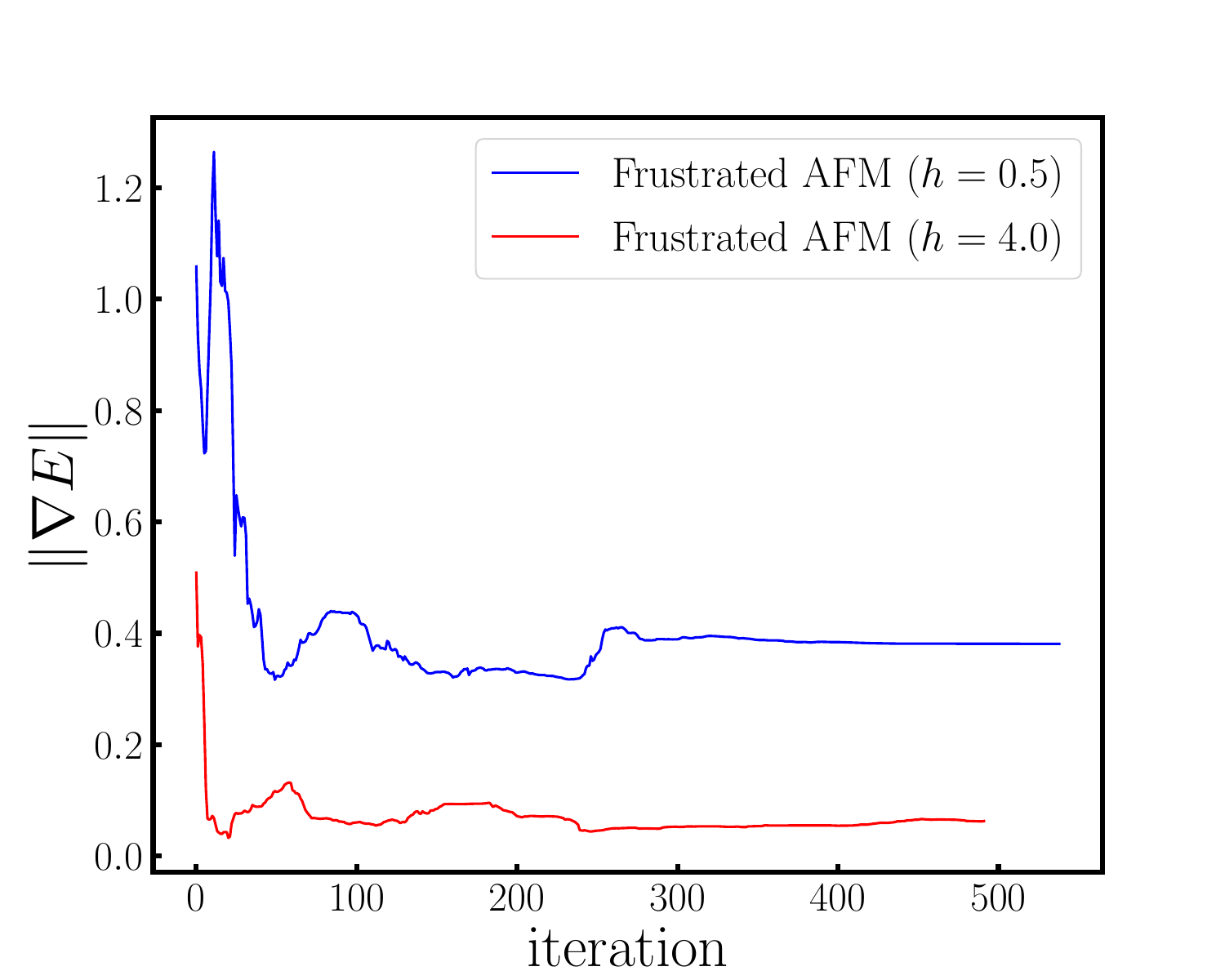}
\includegraphics[width=0.233\textwidth]{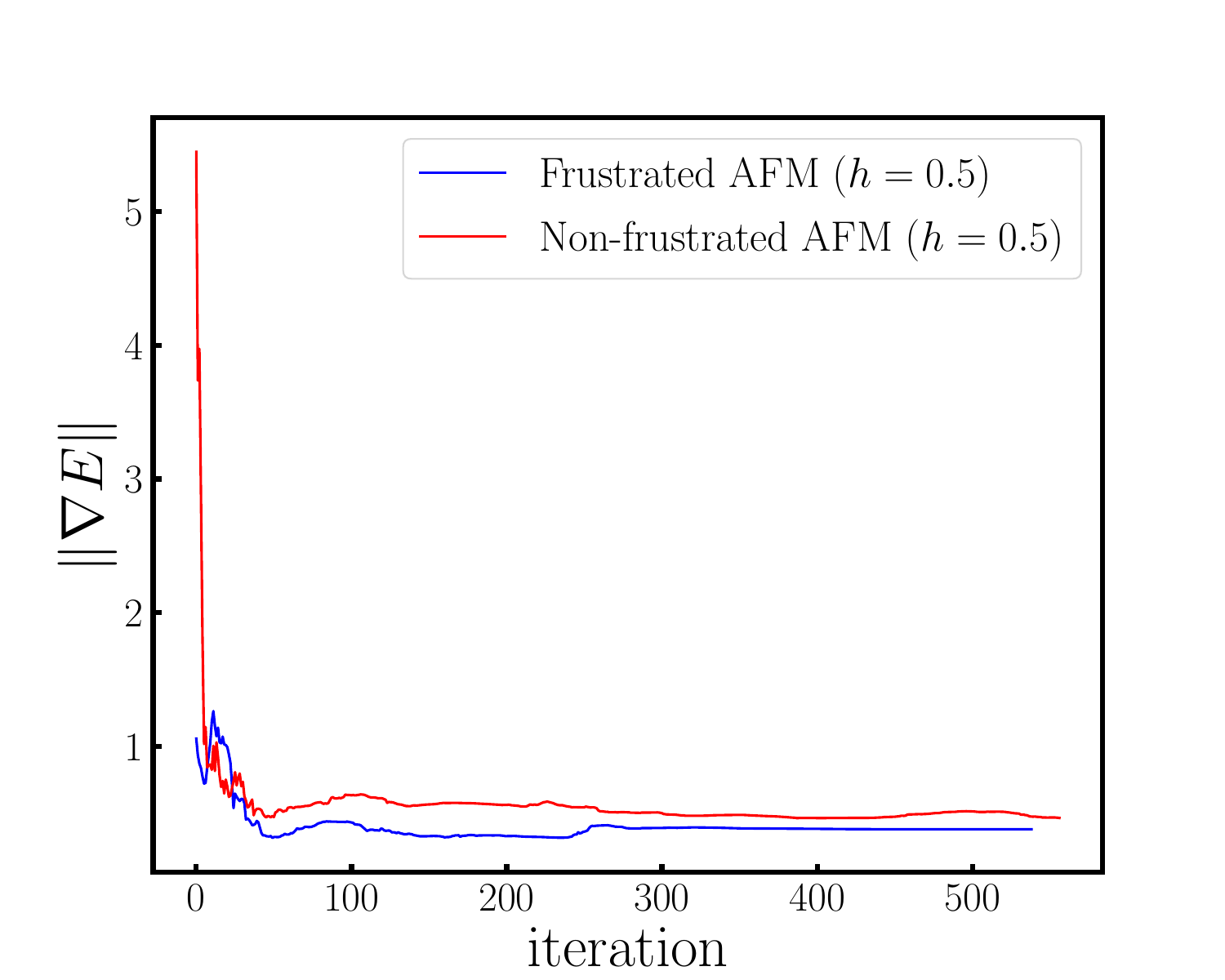}
\caption{Gradient norm versus optimization iteration for the $3\times3$ TFIM using the HVA at fixed depth $p=12$.}
\label{fig:grad_square_triangle}
\end{figure}

To further diagnose the origin of the observed performance limitations, we analyze the gradient norm during optimization at fixed circuit depth $p=12$, shown in Fig.~\ref{fig:grad_square_triangle}.

In the large-field regime ($h=4.0$), the gradient norm decays smoothly and the variational state reaches near-unit fidelity, indicating efficient convergence within the ansatz manifold. In contrast, in the frustrated regime ($h=0.5$), the gradient saturates at a finite value while the final fidelity remains significantly lower, despite a comparable number of optimization iterations.

To isolate the role of geometric frustration, the right panel compares frustrated and non-frustrated AFM square lattices at fixed coupling $h=0.5$. Both systems exhibit comparable, non-vanishing gradient magnitudes, indicating that the optimization landscape remains well-conditioned. However, while the non-frustrated system converges to a high-fidelity state, the frustrated system remains trapped in a suboptimal solution.

The persistence of finite gradients in the frustrated case demonstrates that the observed degradation in performance is not associated with barren plateaus, but instead reflects intrinsic expressibility limitations of the ansatz.

Finally, we examine larger system sizes using Krylov subspace diagonalization as a reference (see Appendix~\ref{app:skqd-res}). The convergence behavior indicates that, in the paramagnetic regime, a relatively small Krylov subspace captures the low-energy physics, whereas in the frustrated regime significantly larger subspaces are required, reflecting increased spectral complexity. For the results presented here, we use a Krylov subspace dimension $\mathcal{D}=96$, corresponding to the largest subspace considered, and assess convergence across different regimes in Appendix~\ref{app:skqd-res}.

\begin{figure}
\centering
\includegraphics[width=0.45\textwidth]{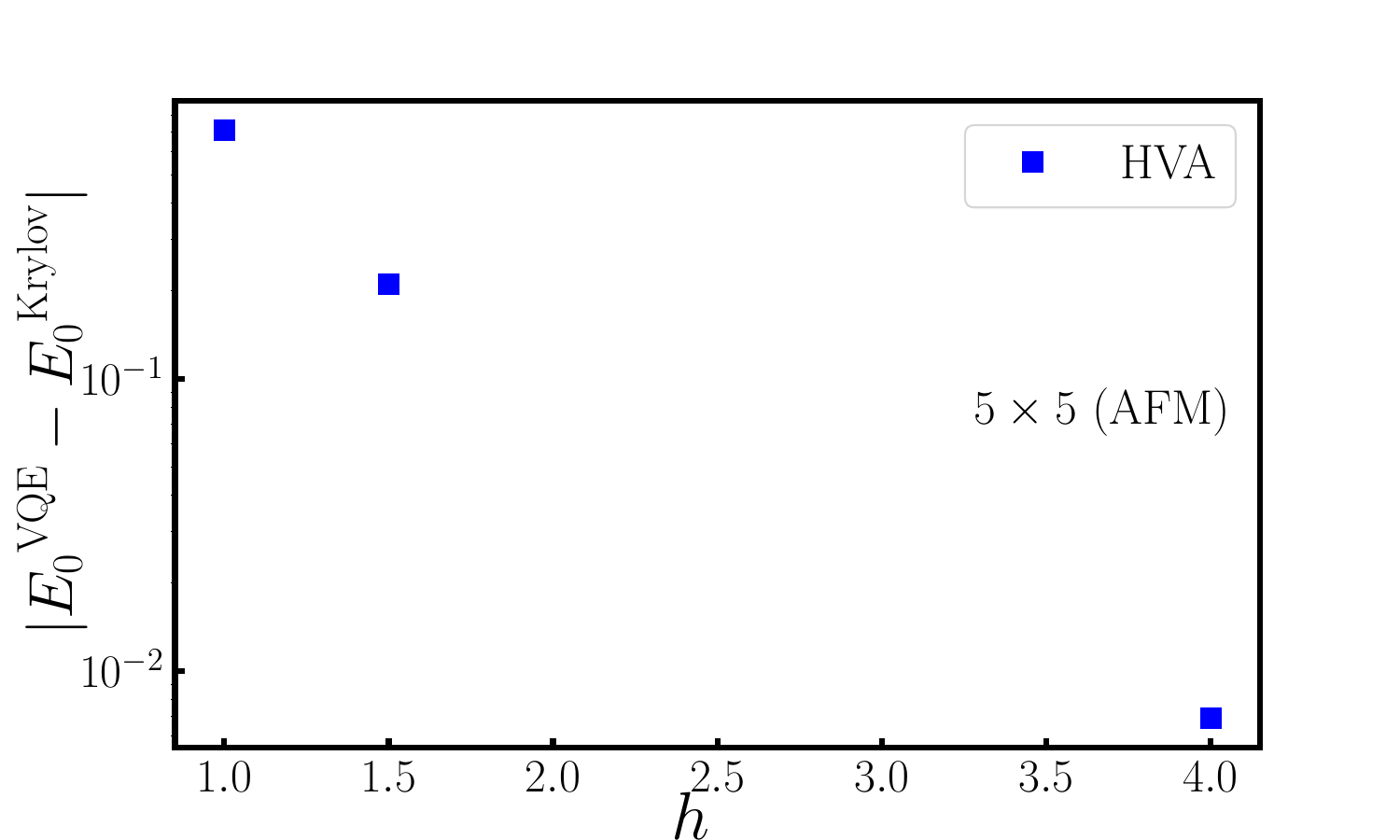}
\caption{Absolute energy error $|E_0^{\mathrm{VQE}} - E_0^{\mathrm{Krylov}}|$ for the frustrated AFM on $5\times5$ lattice at $h=1.0, 1.5$ and $h=4.0$. We used HVA to compute $E_0$ at fixed circuit depth $p=4$.}
\label{fig:5x5-E0-err}
\end{figure}

For the $5\times5$ lattice, the number of interaction terms grows significantly ($|\mathcal{E}|=56$), making the bond-resolved ansatz computationally demanding due to the large parameter space. We therefore restrict the analysis to the standard HVA at fixed circuit depth $p=4$ and a finite optimization budget. While full convergence is not achieved, the comparison is performed under identical conditions for both field values. We observe that the energy error remains small in the paramagnetic regime ($h=4.0$), whereas a noticeably larger deviation appears in the intermediate-field regime $(h = 1.0–1.5)$, indicating increased difficulty near the crossover region. This behavior is consistent with the trends observed for smaller systems and supports the conclusion that frustration-induced expressibility limitations persist beyond system sizes accessible to ED. The slow convergence and enhanced spectral complexity observed in the frustrated regime are consistent with the presence of a highly constrained low-energy manifold arising from competing interactions.

\section{Conclusion}
\label{conclusion}

In this work, we have systematically investigated the performance of variational quantum algorithms in a geometrically frustrated setting using the antiferromagnetic TFIM on a square lattice with frustrated diagonal couplings. By benchmarking against ED and Krylov-based methods, we established a direct connection between geometric frustration, correlation structure, and variational expressibility.

We find that in the frustrated regime, the ground state exhibits strongly inhomogeneous bond correlations that cannot be efficiently captured by Hamiltonian-inspired ansätze with globally shared parameters. As a consequence, the standard HVA exhibits a rapid growth in the circuit depth required to reach high fidelity, along with clear saturation of performance at fixed depth.

Through controlled comparisons with non-frustrated systems and gradient-norm analysis, we demonstrate that this degradation is not solely explained by spectral gap suppression or barren plateaus, but instead originates from an intrinsic expressibility limitation of the ansatz. Geometric frustration induces heterogeneous correlation patterns that cannot be represented within a uniform variational structure, leading to convergence at finite gradients and reduced fidelity.

We show that these limitations can be overcome by introducing a bond-resolved HVA, in which independent parameters are assigned to each interaction term. This approach restores expressibility by adapting to local correlation structure, enabling accurate ground-state preparation at significantly reduced circuit depth.

Extending beyond ground-state properties, we investigated low-energy excitations using both symmetry-resolved constructions and VQD methods. While excitation gaps are accurately captured in regimes with well-separated spectra, near-degenerate manifolds in the frustrated regime present intrinsic challenges, highlighting the interplay between spectral structure and variational expressibility.

Finally, by extending our analysis to larger system sizes using Krylov-based benchmarks, we show that these limitations persist beyond system sizes accessible to ED, indicating that the observed behavior reflects an intrinsic feature of frustrated quantum systems rather than a finite-size artifact.

While the bond-resolved HVA significantly improves performance, its parameter count grows with the number of interactions, posing challenges for larger systems. This suggests that further progress will require structured ansätze that retain local flexibility while controlling parameter growth, for example by exploiting underlying correlation patterns to reduce redundancy, which may become increasingly important as system size increases beyond those accessible in this work.

Overall, our results establish a physically grounded mechanism linking geometric frustration to expressibility limitations in variational quantum algorithms. These findings provide general design principles for constructing problem-informed ansätze capable of efficiently representing complex many-body states, and offer guidance for the application of quantum algorithms to frustrated and strongly correlated systems on near-term quantum devices.

\section*{Acknowledgments}

We thank Debasish Banerjee for pointing out related references and for helpful comments on the draft. This work is supported partly by the National Natural Science Foundation of China under Grants No.~12325508, No.~12293064, and No.~12293060. Part of the numerical computations in this work were carried out on the Nuclear Science Computing Center at Central China Normal University (NSC$^{3}$) and Wuhan Supercomputing Center.

\bibliographystyle{apsrev4-2} 
\showtitleinbib
\bibliography{fixapsbib,refs}
\onecolumngrid\vspace*{2\baselineskip}\twocolumngrid

\appendix
\section{Benchmark of Additional Observables}
\label{app:observables}

\begin{figure}[H]
\centering
\includegraphics[width=0.45\textwidth]{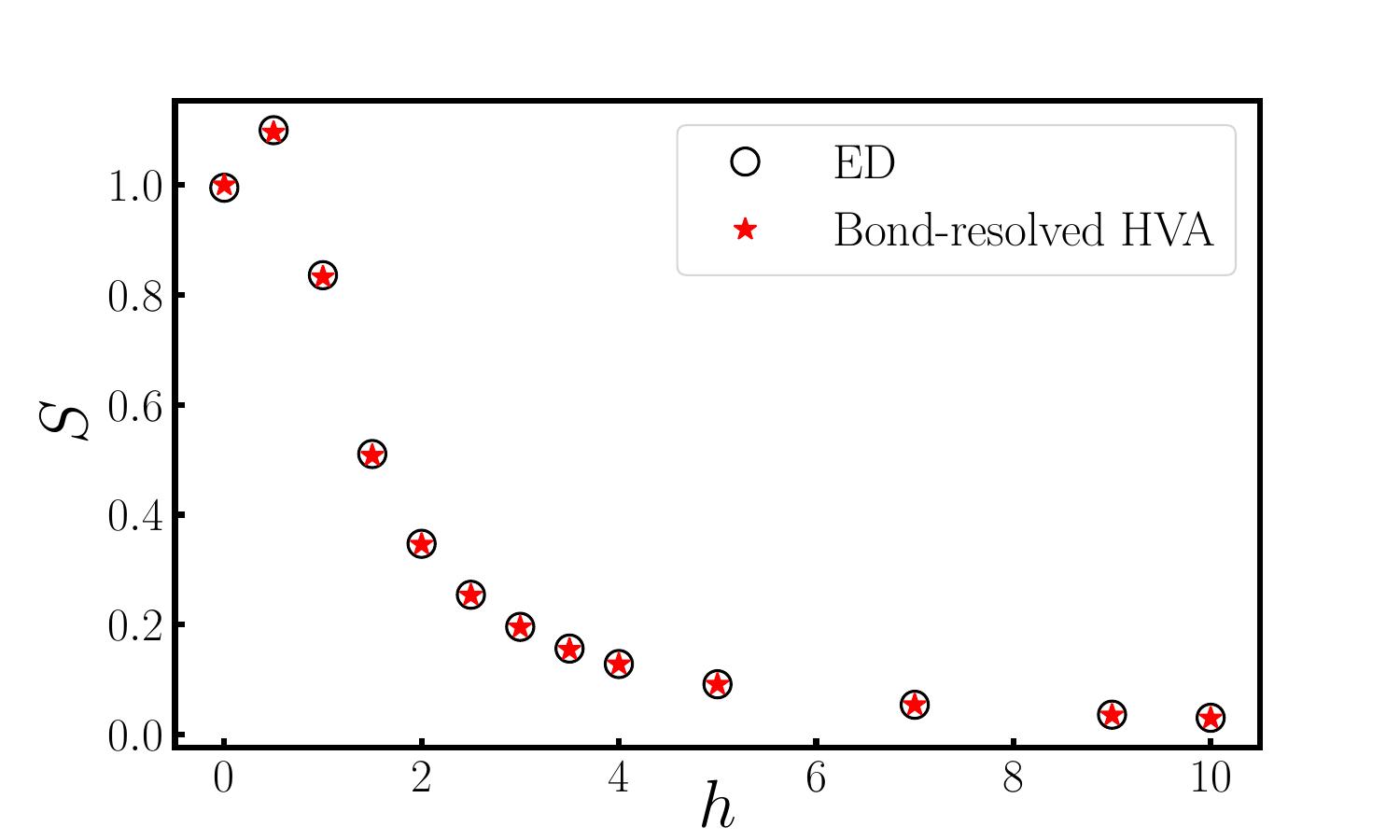}
\includegraphics[width=0.45\textwidth]{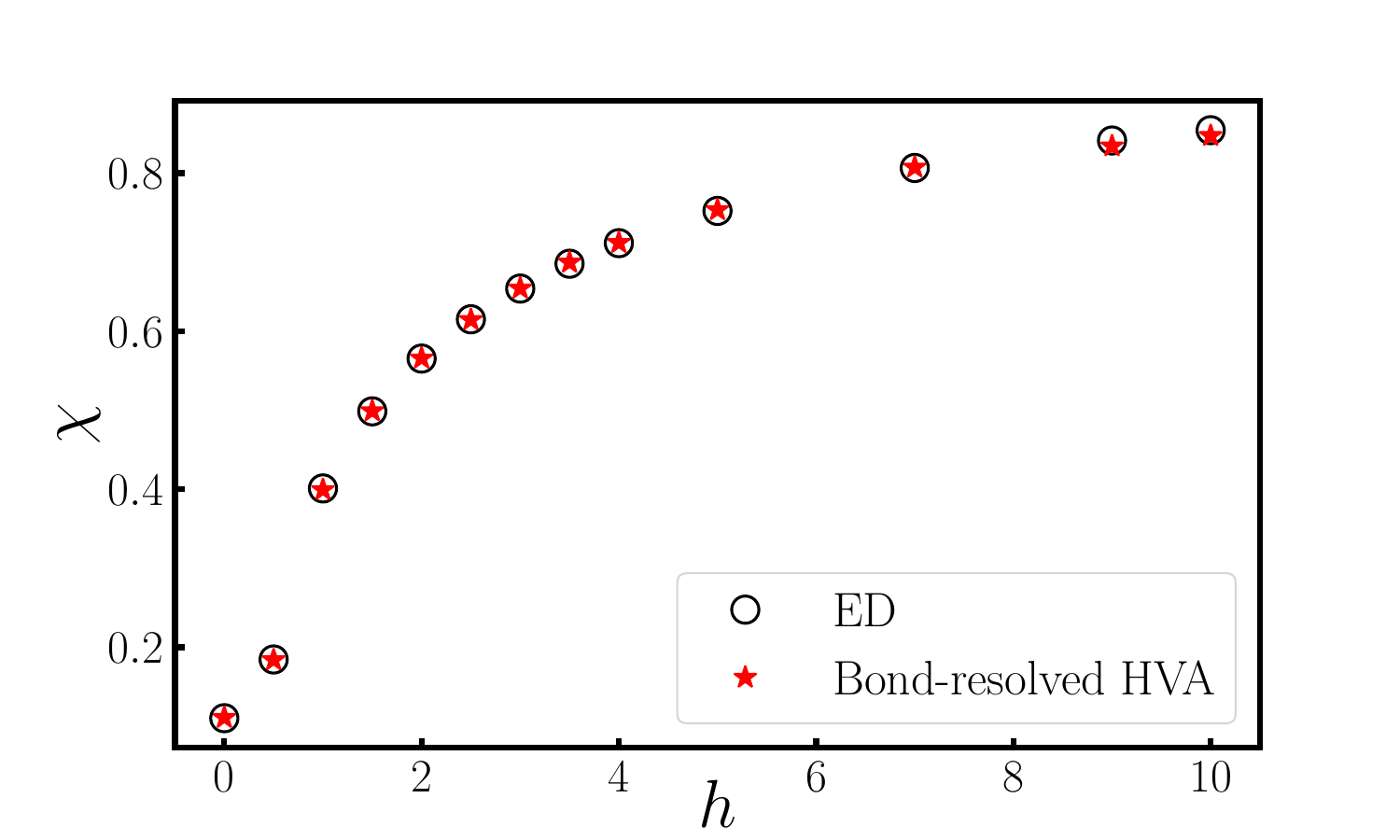}
\caption{(Top) Entanglement entropy ($S$) as a function of the transverse field $h$ for the $3\times3$ frustrated square lattice, comparing ED and VQE using the bond-resolved HVA at fixed depth $p=8$. The ansatz captures the large entanglement in the frustrated low-field regime and its reduction toward the paramagnetic phase. (Bottom) Susceptibility $\chi$ as a function of $h$, showing good agreement between ED and VQE across the full range. The smooth increase of $\chi$ reflects the system's progressive response to $h$.}
\label{fig:susp-ee}
\end{figure}

To further assess the quality of the variational states, we benchmark additional observables obtained from the bond-resolved HVA ($p=8$) against ED for the $3\times 3$ square lattice, focusing on the Entanglement entropy ($S$) and susceptibility $\chi$, shown in Fig.~\ref{fig:susp-ee}.

The variational results show good agreement with ED across the full range of transverse field $h$. In the low-field regime ($h \lesssim 1$), both $S$ and $\chi$ reflect strong correlations associated with the frustrated, near-degenerate manifold. The entanglement entropy is maximal at small $h$ and decreases with increasing field, while the excitation gap begins to open around $h \sim 1$, signaling the lifting of degeneracy. At larger $h$, the system enters a gapped paramagnetic regime, where both observables are smoothly captured by the ansatz.

These results demonstrate that the variational ansatz accurately reproduces not only ground-state energies but also correlation-sensitive observables across distinct physical regimes.

\section{Depth Scaling in Frustrated and Non-Frustrated Systems}
\label{app:depth-scaling}

\begin{figure}[H]
\centering
\includegraphics[width=0.45\textwidth]{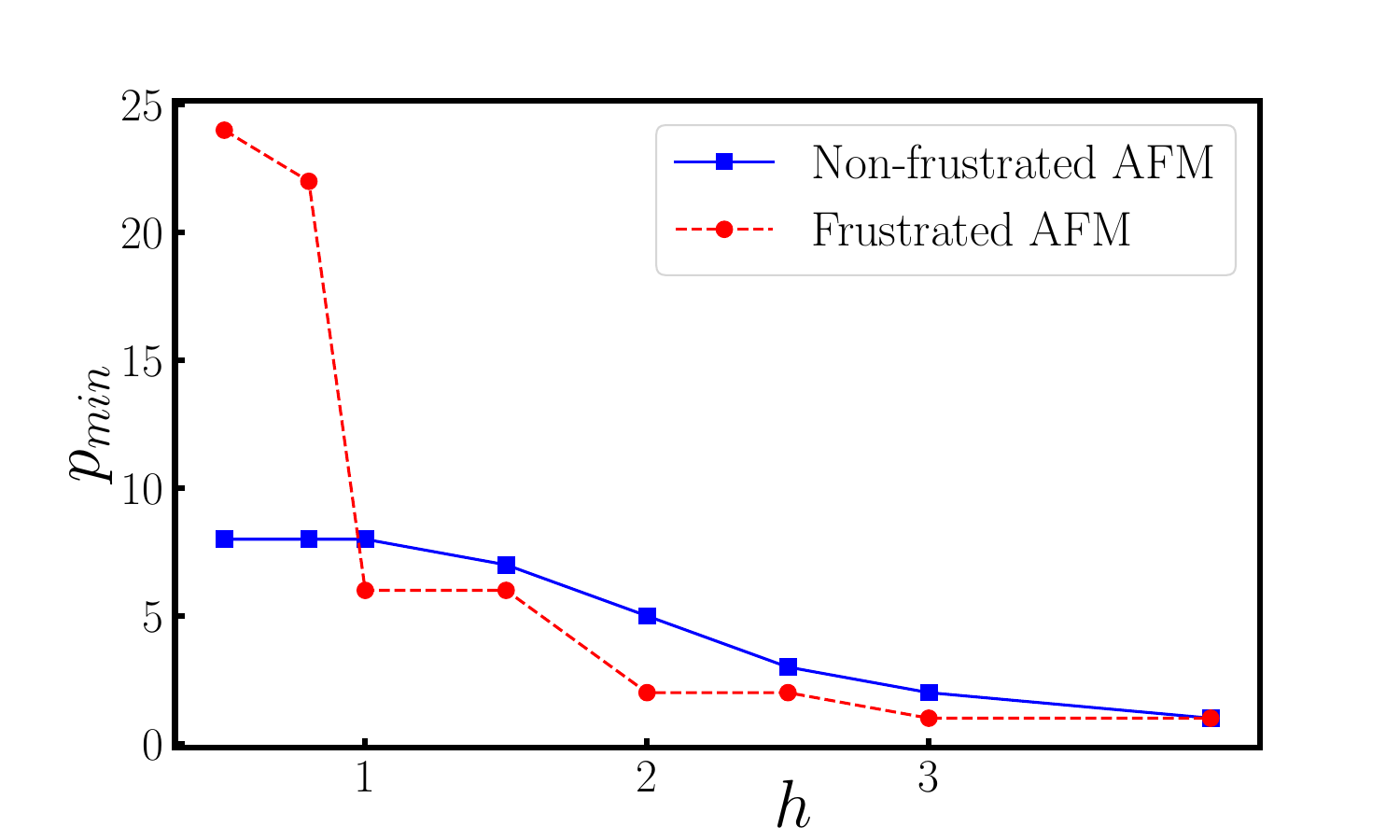}
\caption{Minimum depth $p_{\min}$ required to reach fidelity $f \ge 0.99$ as a function of transverse field $h$ for the HVA on a $3\times3$ AFM square lattice. The frustrated case includes diagonal bonds (triangular plaquettes), while the non-frustrated case includes only nearest-neighbor (horizontal and vertical) bonds.}
\label{fig:depth-scaling}
\end{figure}

To further isolate the role of geometric frustration, we compare the minimum circuit depth $p_{\min}$ required to reach fidelity $f \ge 0.99$ for the HVA on frustrated and non-frustrated AFM lattices at fixed system size ($3\times3$), as shown in Fig.~\ref{fig:depth-scaling}. Here, the frustrated model includes diagonal bonds (triangular plaquettes), while the non-frustrated model includes only horizontal and vertical nearest-neighbor couplings. 

In the non-frustrated case, correlations remain approximately homogeneous across bonds, and the standard HVA is sufficient to accurately capture the ground state. We therefore restrict the comparison to the standard HVA for both models to isolate the effect of frustration-induced correlation inhomogeneity.

The non-frustrated AFM achieves high fidelity with shallow circuits ($p_{\min} \lesssim 8$) across all $h$, whereas the frustrated AFM requires substantially larger depth in the low-field regime (e.g., $p_{\min} \approx 24$ at $h=0.5$).

This disparity demonstrates that the increased depth requirement is not solely due to spectral effects, but arises from frustration-induced correlation complexity.

\section{Optimizer Performance}
\label{app:optimizer}

\begin{figure}[H]
\centering
\includegraphics[width=0.45\textwidth]{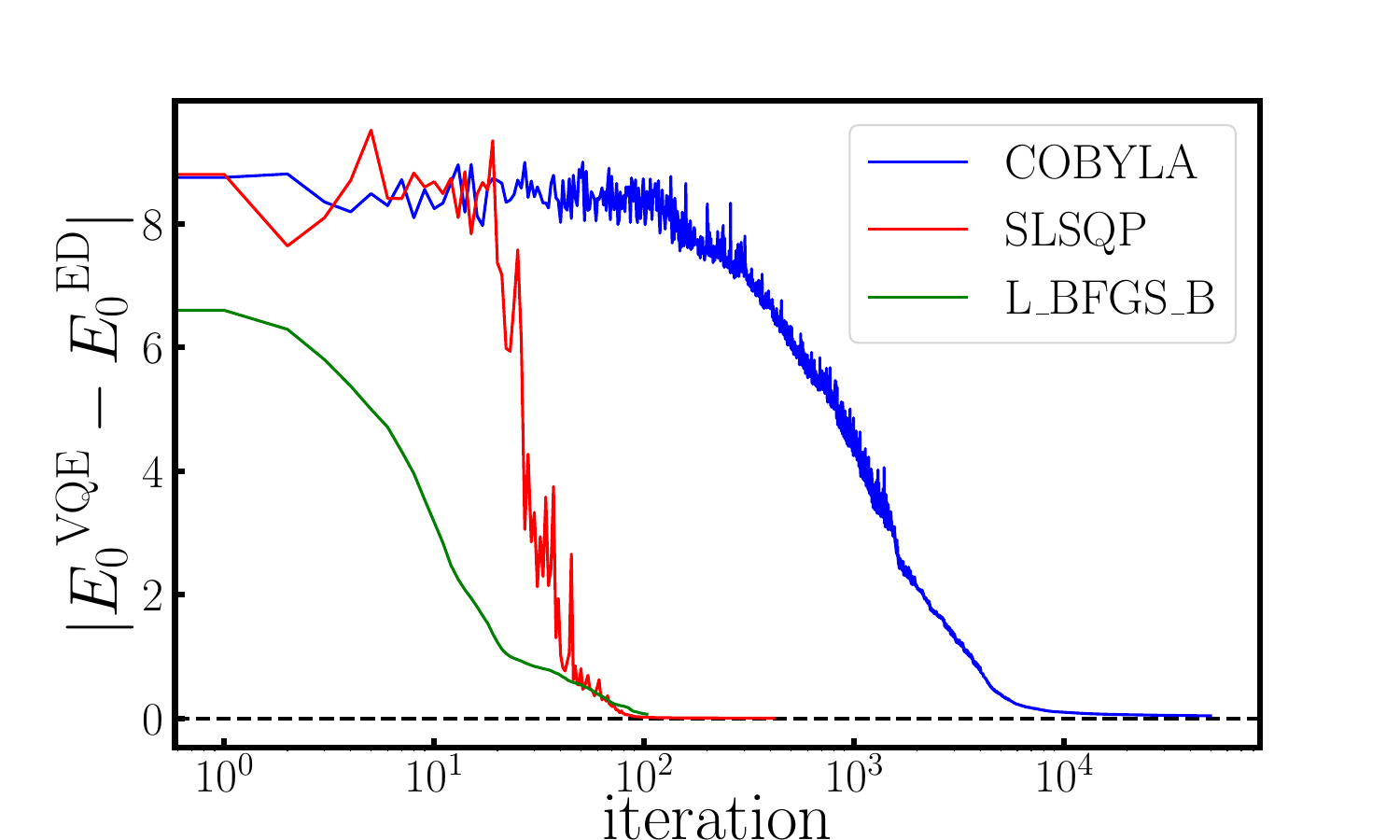}
\caption{Energy error $|{E_0}^{\mathrm{VQE}} - E_0^{\mathrm{ED}}|$ as a function of optimization iteration for different classical optimizers applied to the bond-resolved HVA on a $3\times3$ frustrated AFM lattice at $h=0.5$ with fixed depth $p=8$.}
\label{fig:optimizer}
\end{figure}

We compare the performance of different classical optimizers for variational ground-state preparation using the bond-resolved HVA. Figure~\ref{fig:optimizer} shows the energy error $|E_0^{\mathrm{VQE}} - E_0^{\mathrm{ED}}|$ as a function of optimization iteration for the $3\times3$ frustrated AFM lattice at $h=0.5$ with fixed circuit depth $p=8$, where all optimizers initialized from the same parameter configuration.

We consider COBYLA, SLSQP, and L-BFGS-B. Among these, SLSQP converges most rapidly, reaching near-zero energy error within $\mathcal{O}(10^2)$ iterations and achieving near-unit fidelity ($f > 0.99$). L-BFGS-B also converges efficiently but more gradually, while COBYLA requires significantly more iterations and attains lower final fidelity ($f \approx 0.96$).

These results indicate that gradient-based optimizers are more effective in navigating the variational landscape of the bond-resolved ansatz, particularly in the frustrated regime, where the parameter space is higher-dimensional and the energy landscape is more structured.

\section{Krylov Subspace Convergence Across Physical Regimes}
\label{app:skqd-res}

To further probe the complexity of the low-energy spectrum, we examine the convergence of the ground-state energy $E_0$ using a Krylov subspace (Rayleigh–Ritz) diagonalization approach for the $4\times4$ and $5\times5$ square lattices.

\begin{figure}[H]
\centering
\includegraphics[width=0.45\textwidth]{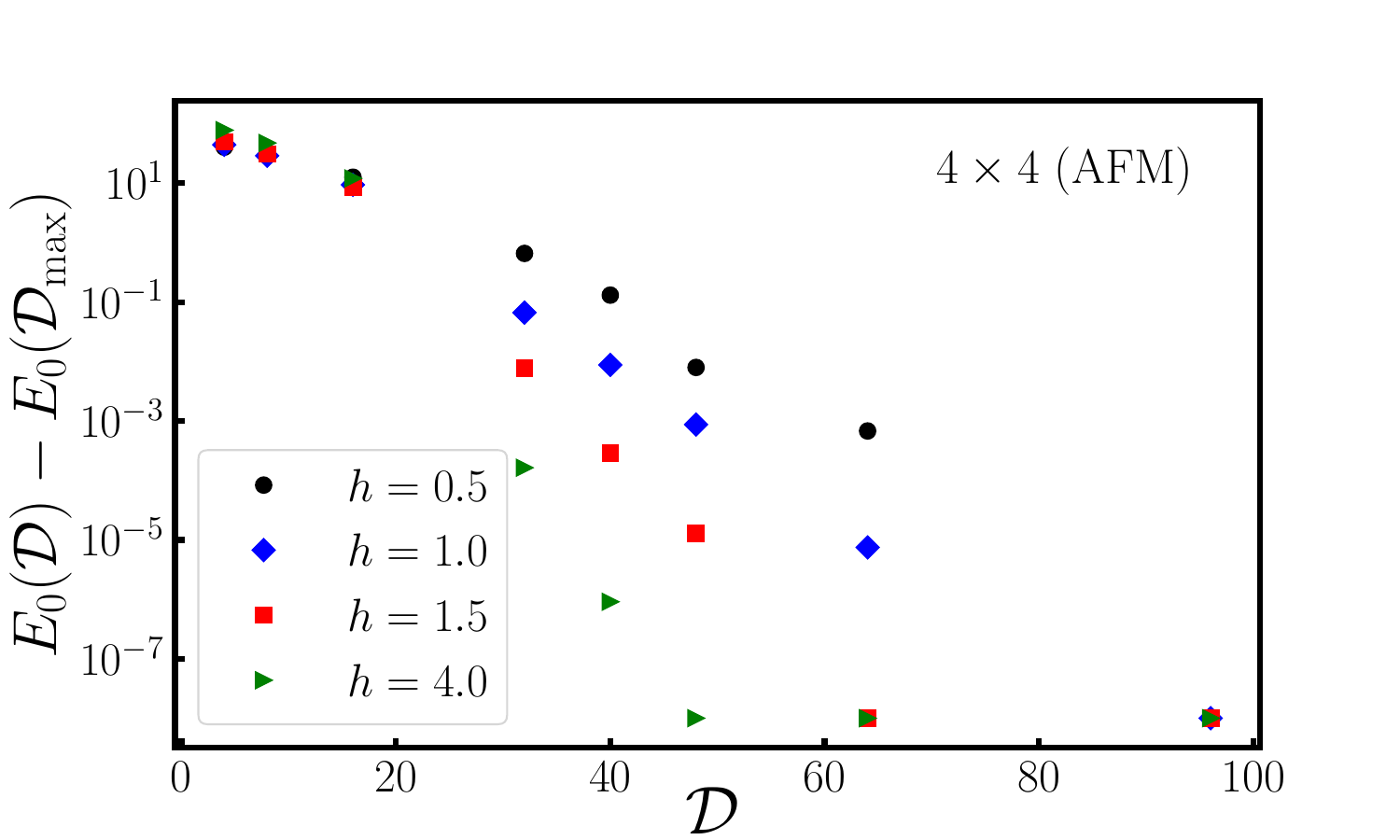}
\includegraphics[width=0.45\textwidth]{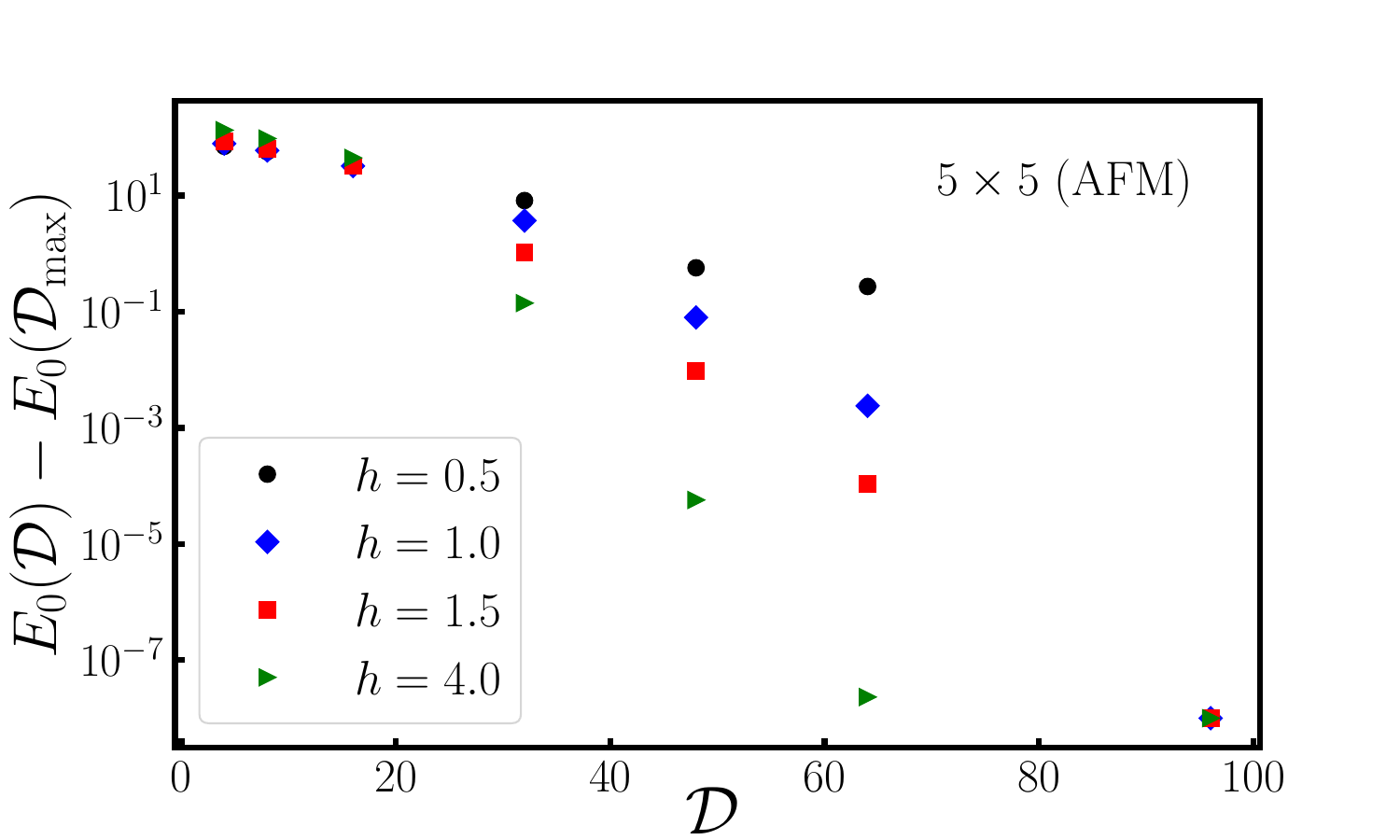}
\caption{Convergence of the ground-state energy with Krylov subspace dimension $\mathcal{D}$ for the $4\times 4$ (Top) and $5\times5$ (Bottom) frustrated lattices at representative transverse fields $h$. The absolute energy difference relative to the largest subspace dimension is shown on a logarithmic scale.}
\label{fig:skqd-conv}
\end{figure}

Figure~\ref{fig:skqd-conv} shows the convergence of the ground-state energy as a function of the Krylov subspace dimension $\mathcal{D}$ for both system sizes and representative values of the transverse field $h$.

For the $4\times4$ lattice, where ED is available, the Krylov estimates at the largest subspace dimension ($\mathcal{D}=96$) show excellent agreement with the exact ground-state energy. The deviation is negligible across the range at $h=1.0$--$4.0$, and remains below $10^{-2}$ even in the most challenging frustrated regime ($h=0.5$).

As the system size increases from $4\times 4$ to $5\times 5$, the convergence with increasing $\mathcal{D}$ becomes slower. This effect is strongest in the frustrated regime (low $h$), where much larger Krylov subspaces are needed. In the paramagnetic regime ($h=4.0$), the $5\times5$ system still converges well, but requires larger $\mathcal{D}$ compared to $4\times4$, indicating a systematic shift in convergence with system size.

These results demonstrate that the increased difficulty observed in variational approaches in the frustrated regime is not an artifact of the ansatz, but instead reflects the intrinsic complexity of the underlying many-body spectrum.

\end{document}